\documentclass[a4paper,11pt]{article}
\usepackage{amsmath,amssymb,color,comment}
\usepackage{slashed, tensor, bm, physics}
\usepackage{caption}
\usepackage{graphicx}
\usepackage{multirow}
\usepackage{float}
\usepackage[compat=1.1.0]{tikz-feynhand}
\usepackage{jheppub} 

\usepackage[T1]{fontenc} 
\usepackage{booktabs} 

\newcommand{\U}{{\rm U}}

\title{Axionic Wormholes in Metric-Affine Gravity}
\author{Shonosuke Takeshita}
\author{and Naoki Yoshioka}

\affiliation{Physics Program, Graduate School of Advanced Science and Engineering, Hiroshima University, Higashi-Hiroshima 739-8526, Japan}

\emailAdd{shonosuke@hiroshima-u.ac.jp}
\emailAdd{na-yoshioka@hiroshima-u.ac.jp}

\abstract{
The axion is a promising candidate for solving the strong CP problem.
To solve this problem, the global $\U(1)$ symmetry must be preserved to a high degree of accuracy.
However, it is well known that global symmetries are explicitly violated by quantum gravity effects, giving rise to what is referred to as the axion quality problem.
In this paper, we investigate axionic wormholes as a source of explicit $\U(1)$ violation in Metric-Affine Gravity.
This framework allows for spacetime torsion and non-metricity, which accommodate additional curvature-like and topological terms, such as the Holst and Nieh--Yan terms, that are absent from the metric and Palatini formalisms.
We show that non-minimal couplings to these terms modify the wormhole dynamics and enhance the Euclidean wormhole action, thereby alleviating the axion quality problem. 
We also find that the viable parameter space is enlarged when two of these couplings are simultaneously present.
We further identify representative parameter regions where the alleviation of the axion quality problem is compatible with inflationary constraints.
}

\begin{document} 
\begin{flushright}
HUPD-2602
\end{flushright}
\maketitle
\flushbottom

\section{Introduction}
\label{sec:Intro}

The axion~\cite{Wilczek:1977pj, Weinberg:1977ma, Kim:2008hd, Graham:2015ouw}, associated with the spontaneous breaking of the global $\U(1)$ Peccei--Quinn (PQ) symmetry~\cite{Peccei:1977hh, Peccei:1977ur}, offers a compelling solution to the strong CP problem~\cite{Baluni:1978rf, Crewther:1979pi}.
The axion obtains a periodic potential through non-perturbative QCD effects and dynamically relaxes to a CP-conserving vacuum.
For this mechanism to work, the PQ symmetry must be protected to an extremely high degree to prevent the potential minimum from deviating from the CP-conserving point.
However, it is well known that quantum gravity does not allow exact global symmetries~\cite{Banks:2010zn, Witten:2017hdv, Harlow:2018jwu, Harlow:2018tng}.
This implies the existence of explicit PQ symmetry breaking induced by quantum gravity, which may spoil the axion solution.
In fact, such breaking effects have been investigated using the effective field theory approach~\cite{Georgi:1981pu, Dine:1986bg, Ghigna:1992iv}, and it has been found that, generically, they are too large to maintain the axion solution to the strong CP problem.
This is referred to as the axion quality problem~\cite{Kamionkowski:1992mf, Holman:1992us, Barr:1992qq}.

Axionic wormholes~\cite{Giddings:1987cg, Lee:1988ge, Coleman:1989zu, Kallosh:1995hi, Hebecker:2018ofv}, which are gravitational instantons in Euclidean spacetime, are one of the sources that explicitly induce the violation of the global $\U(1)$ symmetry by gravity.
Geometrically, these configurations connect two asymptotically flat regions via a throat and carry a conserved PQ charge.
Due to the flow of PQ charge through these wormholes, explicit PQ violation occurs in our universe.
This breaking effect is described by effective local operators at low energies~\cite{Giddings:1988cx, Coleman:1988cy, Rey:1989mg, Abbott:1989jw}.
The contributions of these operators are exponentially suppressed by the factor $e^{-S}$, where $S$ denotes the wormhole action.
To maintain the axion solution to the strong CP problem, the wormhole action is required to be sufficiently large, typically $S \gtrsim 190$~\cite{Kallosh:1995hi, Alvey:2020nyh}.
In Ref.~\cite{Giddings:1987cg}, Giddings and Strominger derived the axionic wormhole solution in a setup where the radial mode of the complex scalar is fixed at its vacuum expectation value, $f(x)=f_a$.
In this case, the wormhole action scales as $S \sim M_{p} / f_a$, and for a sufficiently small axion decay constant, $f_a \lesssim 10^{16}$~GeV~\cite{Alonso:2017avz}, the effect of explicit symmetry breaking is suppressed enough to avoid the axion quality problem.
However, once the radial mode is treated dynamically, the action is significantly reduced to a logarithmic form, $S \sim \log(M_p/f_a)$~\cite{Kallosh:1995hi}.
Consequently, gravitational violations of the PQ symmetry are no longer suppressed, and the axion quality problem persists in general relativity~(GR)~\cite{Kallosh:1995hi,Alonso:2017avz,Alvey:2020nyh}.

Modified gravitational theories, considered as extensions of GR, offer a promising solution to the axion quality problem.
Ref.~\cite{Hamaguchi:2021mmt} discusses the case where the PQ scalar field has a non-minimal coupling to gravity in the metric formalism.
The authors demonstrated that the non-minimal coupling can alleviate the axion quality problem while remaining consistent with inflation.
A similar approach using the Palatini formalism was explored in Ref.~\cite{Cheong:2022ikv}, which addresses the axion quality problem and inflation simultaneously.
For background on the Palatini formalism and inflation, see Refs.~\cite{Ferraris:1982wci, Sotiriou:2008rp, Olmo:2011uz} and \cite{Guth:1980zm, Linde:1981mu, Linde:1983gd, Olive:1989nu, Baumann:2009ds}, respectively.
Axionic wormholes beyond GR have also been investigated in Refs.~\cite{Andriolo:2022rxc, Cheong:2023hrj, Kanazawa:2023xzy, Cheong:2024kum}.

In this paper, we discuss axionic wormholes in Metric-Affine Gravity~(MAG)~\cite{Hehl:1994ue, Blagojevic:2012bc} with a non-minimally coupled scalar field.
Unlike in GR, the metric $g_{\mu\nu}$ and the affine connection ${\Gamma^{\mu}}_{\nu\rho}$ in MAG are treated as independent variables, and thus the affine connection is not determined solely by the dynamics of spacetime.
Due to the presence of spacetime torsion and non-metricity, we can consider additional curvature-like and topological scalar quantities, specifically the Holst term~\cite{Nelson:1980ph, Holst:1995pc, Barbero:1994ap, Immirzi:1996di} and the Nieh--Yan term~\cite{Nieh:1981ww}, which are absent in GR and the Palatini formalism.
The Holst term has historically attracted attention in the context of the Hamiltonian formulations of gravity and loop quantum gravity~\cite{Rovelli:1997yv,Ashtekar:2004eh}.
Its parity-odd nature and couplings to the fermion sector have also motivated phenomenological and cosmological studies~\cite{Freidel:2005sn,Mercuri:2006um,deBerredo-Peixoto:2012psm}.
On the other hand, the Nieh--Yan term is a topological invariant in the Einstein--Cartan theory and has been discussed in relation to the chiral anomaly in spacetime with torsion~\cite{Chandia:1997hu}.
Recently, modified gravitational theories involving these terms have also been investigated in cosmological and phenomenological contexts~\cite{Gialamas:2022xtt, He:2023vlj, Gialamas:2023emn, Gialamas:2024uar, Gialamas:2024iyu, Salvio:2025izr, Racioppi:2025igu, Katsoulas:2025srh, Gialamas:2026pjo, Dimopoulos:2026iwq}.

We demonstrate that these terms significantly contribute to the axionic wormhole action, thereby offering a solution to the axion quality problem.
Furthermore, we discuss whether the parameter regions required to solve the axion quality problem are compatible with inflationary scenarios that arise in this class of models.

The paper is organized as follows.
Section~\ref{sec:Model} reviews MAG and derives the equations of motion (EoMs) for the metric, the affine connection, and a scalar field in Euclidean spacetime.
In Section~\ref{sec:NA}, we present numerical results concerning the behavior of the axionic wormhole and the dependence of the wormhole action on the non-minimal coupling parameters.
We also discuss the compatibility of the identified parameter spaces with observational constraints on inflation.
Section~\ref{sec:Conclusion} is devoted to conclusions and discussion.
In Appendix~\ref{sec:F0A0}, we present the initial values of the wormhole solutions and compare their respective features.
In Appendix~\ref{sec:inflation}, we summarize the inflationary analysis.

Calculations in this paper are based on the following notation.
$M_p = \sqrt{1/8\pi G_N}$ is the reduced Planck mass.
$G_N$ denotes Newton's gravitational constant.
The Minkowski metric $\eta_{\mu\nu}$ is defined by $\eta=\text{diag}(-,+,+,+)$.
The totally antisymmetric symbol $\epsilon^{\mu\nu\rho\sigma}$ takes the value $\epsilon^{0123}\equiv 1$.
The completely antisymmetric tensor $\varepsilon^{\mu\nu\rho\sigma}$ is defined by $\varepsilon^{\mu\nu\rho\sigma} = \epsilon^{\mu\nu\rho\sigma} / \sqrt{|\text{det}(g_{\mu\nu})|}$.
The symmetrization $A_{\lbrace \mu\nu\rbrace}$ and the antisymmetrization $A_{[\mu\nu]}$ are respectively $\frac{1}{2}(A_{\mu\nu} + A_{\nu\mu})$ and $\frac{1}{2}(A_{\mu\nu} - A_{\nu\mu})$.
Greek indices denote spacetime coordinate indices.

\section{Model}
\label{sec:Model}
In MAG, the metric $g_{\mu\nu}$ and the affine connection ${\Gamma^{\mu}}_{\nu\rho}$ are treated as independent variables.
The metric defines the spacetime distance via $ds^2 = g_{\mu\nu}dx^{\mu}dx^{\nu}$, while the affine connection determines parallel transport and defines the covariant derivative $\nabla$ as $\nabla_{\mu}A^{\nu}\equiv \partial_{\mu}A^{\nu}+{\Gamma^{\nu}}_{\alpha\mu}A^{\alpha}$.
Since no condition is imposed on these variables in this theory, the affine connection generally has an antisymmetric part (torsion) ${T^{\mu}}_{\nu\rho}$ and non-metricity $Q_{\mu\nu\rho}$.
They are defined by
\begin{align}
    {T^{\mu}}_{\nu\rho} &= {\Gamma^{\mu}}_{\nu\rho} - {\Gamma^{\mu}}_{\rho\nu}, \\
    Q_{\mu\nu\rho} &= \nabla_{\mu}g_{\nu\rho}.\label{eq:non-metricity}
\end{align}
Using these definitions, the affine connection $\Gamma$ can be separated into the Levi-Civita connection ${\Gamma^{\diamond\mu}}_{\nu\rho}$ and distortion ${N^{\mu}}_{\nu\rho}$ parts,
\begin{align}
{\Gamma^{\mu}}_{\nu\rho} &= {\Gamma^{\diamond\mu}}_{\nu\rho} + {N^{\mu}}_{\nu\rho}, \\
{\Gamma^{\diamond\mu}}_{\nu\rho} &= \frac{g^{\mu\sigma}}{2}(g_{\nu\sigma,\rho} + g_{\rho\sigma,\nu} - g_{\nu\rho,\sigma}), \\
{N^{\mu}}_{\nu\rho} &= \frac{1}{2}({T^{\mu}}_{\nu\rho} + {T_{\nu\rho}}^{\mu} + {T_{\rho\nu}}^{\mu} + {Q^{\mu}}_{\nu\rho} - {Q_{\nu\rho}}^{\mu} - {Q_{\rho\nu}}^{\mu}).
\end{align}
Similarly, the curvature scalar can be written as 
\begin{align}
  R =R^{\diamond}+ Q + T + T_{\mu}(Q^{\mu} - \hat{Q}^{\mu}) - T_{\mu\nu\rho}Q^{\nu\rho\mu} + \nabla^{\diamond}_{\mu}B^{\mu},
\end{align}
with 
\begin{align}
    Q &= -\frac{1}{4}(2Q^{\mu\nu\rho}Q_{\rho\mu\nu} - Q^{\mu\nu\rho}Q_{\mu\nu\rho}) + \frac{1}{2}Q_{\mu}(\hat{Q}^{\mu} - \frac{1}{2}Q^{\mu}),\\   
    T 
    &= \frac{1}{4} T^{\rho\mu\nu} T_{\rho\mu\nu}
    - \frac{1}{2} T^{\rho\mu\nu} T_{\mu\nu\rho}
    - T^\mu T_\mu,\\
    T_{\mu} &= {T^{\nu}}_{\mu\nu},\quad Q_{\mu} = {Q_{\mu\nu}}^{\nu}, \quad \hat{Q}_{\mu} = {Q^{\nu}}_{\nu\mu},
\end{align}
and
\begin{align}
   B^{\mu} = Q^{\mu} - \hat{Q}^{\mu} - 2 T^{\mu},
\end{align}
where the superscript $\diamond$ denotes quantities constructed from the Levi-Civita connection ${\Gamma^{\diamond\mu}}_{\nu\rho}$.
The Palatini formalism is obtained by assuming the absence of torsion, as required by the equivalence principle.
On the other hand, the Einstein--Cartan theory is obtained by imposing the metric-compatibility condition, $Q_{\mu\nu\rho}=0$, which ensures that vector norms are preserved under parallel transport.
By varying the action with respect to the affine connection ${\Gamma^{\mu}}_{\nu\rho}$, we algebraically determine the distortion parts.

\subsection{Non-minimal couplings in MAG}

In our study, we evaluate the action for the wormhole solution in the presence of a complex scalar field $\Phi$ non-minimally coupled to the curvature scalar $R$, the Holst term $\epsilon R \equiv \epsilon^{\mu\nu\rho\sigma}R_{\mu\nu\rho\sigma}/\sqrt{-g}$ \cite{Nelson:1980ph, Hojman:1980kv, Holst:1995pc, Barbero:1994ap, Immirzi:1996di}, and the Nieh--Yan term $\nabla^{\diamond}_{\mu}\hat{T}^{\mu}=\partial_{\mu}(\epsilon^{\mu\nu\rho\sigma}T_{\nu\rho\sigma})/\sqrt{-g}$ \cite{Nieh:1981ww, Chandia:1997hu}, where $g=\det(g_{\mu\nu})$.
These last two terms, which vanish in torsionless theories such as GR and the Palatini formalism, survive in our model and give rise to qualitatively new effects.
We start with the following action,
\begin{align}
  S = \int \sqrt{-g}d^4x \Big\{
    \frac{F}{2}R  + \frac{H}{4}\epsilon R - \frac{Y}{4}\nabla^{\diamond}_{\mu}\hat{T}^{\mu}
    + \mathcal{L}_{\text{m}}
    \Big\}, \label{eq:original_action}
\end{align}
with
\begin{align}
    F& \equiv M^2_p + \xi(2|\Phi|^2-f^2_a)=M^2_p + \xi(f^2-f^2_a),\label{eq:NM1}\\
    H& \equiv \frac{M^2_p + \xi_H(2|\Phi|^2-f^2_a)}{\gamma_{H}}=\frac{M^2_p + \xi_H(f^2-f^2_a)}{\gamma_{H}},\label{eq:NM2}\\
    Y& \equiv \frac{M^2_p + \xi_{\eta}(2|\Phi|^2-f^2_a)}{\gamma_{\eta}}=\frac{M^2_p + \xi_{\eta}(f^2-f^2_a)}{\gamma_{\eta}},\label{eq:NM3}
\end{align}
where $\xi$, $\xi_H$, and $\xi_{\eta}$ are the non-minimal coupling parameters, $\gamma_{H}$ is the Barbero--Immirzi parameter, $\gamma_{\eta}$ is a dimensionless parameter, and $f_a$ is the axion decay constant.
In the wormhole analysis below, we impose $M_p^2-\xi f_a^2 \geq 0$ so that the effective gravitational constant has the correct sign for arbitrary values of the field $\Phi$.
Owing to the symmetry of the system, these non-minimal coupling functions in Eqs.~\eqref{eq:NM1}--\eqref{eq:NM3} depend only on the radial component $f$ of the complex scalar field $\Phi = f e^{i\theta}/\sqrt{2}$, with $\theta$ denoting the phase.
$\mathcal{L}_{\text{m}}$ is the Lagrangian density of $\Phi$,
\begin{align}
    \mathcal{L}_{\text{m}}= - |\partial \Phi|^2 - \frac{\lambda}{4}(2|\Phi|^2 - f^2_a)^2,
\end{align}
where $\lambda$ is the self-coupling constant.

\subsection{EoMs and action in Euclidean spacetime}
For the analysis of the wormhole solution in a spherically symmetric spacetime, we consider the following metric,
\begin{align}
    ds^2 = dr^2 + a(r)^2(d\varphi^2 + \sin^2\varphi(d\Theta^2 + \sin^2\Theta d\phi^2)).\label{eq:metric}
\end{align}
Here we perform the analytic continuation from the Lorentzian spacetime to the Euclidean spacetime, and hereafter, we label Euclidean quantities with $E$.
In the Euclidean spacetime \eqref{eq:metric}, the action in Eq.~\eqref{eq:original_action} is rewritten as
\begin{eqnarray}
   S^E = -\int \sqrt{g}d^4x_E \Big\{
        \frac{F}{2}R + \frac{iH}{4}\epsilon_E R -\frac{iY}{4}\nabla^{\diamond}_{\mu}\hat{T}^{\mu} - \mathcal{L}^E_{\text{m}}
        \Big\}, \label{eq:euclidean_action}
\end{eqnarray}
with
\begin{align}
    \mathcal{L}^E_{\text{m}} = \frac{1}{2}\partial_{\mu}f\partial^{\mu}f - \frac{1}{2}f^2\partial_{\mu}\theta\partial^{\mu}\theta + V, \label{eq:matter_lagrangian}
\end{align}
where the potential is $V=\frac{\lambda}{4}(f^2 - f^2_a)^2$.
$\epsilon_E$ is the totally antisymmetric symbol $\epsilon^{0123}_E \equiv 1$ in Euclidean spacetime.
In the Lagrangian density in Eq.~\eqref{eq:matter_lagrangian}, we have included the term $\theta\partial_{\mu}(\sqrt{g}f^2\partial^{\mu}\theta)$ to obtain the correct solution in the presence of the conserved quantity $n$,
\begin{align}
    n &\equiv \int d\Omega_3 (\sqrt{g}f^2\partial^{0}\theta) = 2\pi^2 a^3 f^2 \frac{d\theta}{dr}, \label{eq:PQ_charge}
\end{align}
associated with the invariance under shifts of the $\theta$ component~\cite{Lee:1988ge, Coleman:1989zu}.
Therefore, the sign of the second term in Eq.~\eqref{eq:matter_lagrangian} has been flipped.

By taking the variation of the action given in Eq.~\eqref{eq:euclidean_action} with respect to the metric and the affine connection, we derive the Einstein and Cartan equations as follows:
\begin{align}
    R_{\mu\nu} - \frac{1}{2}Rg_{\mu\nu} - i\frac{H}{2F}\varepsilon_E^{\alpha\beta\gamma\delta}g_{\alpha\lbrace\mu}R_{\nu\rbrace\beta\gamma\delta} + i\frac{Y_{f}}{2F}\varepsilon_E^{\alpha\beta\gamma\delta}g_{\alpha\lbrace\mu}T_{\nu\rbrace\beta\gamma}\partial_{\delta}f = -\frac{2}{F}\mathcal{T}_{\mu\nu}, \label{eq:Einstein_equation}
\end{align}
\begin{eqnarray}
    T_{\nu\mu\rho} + 2T_{[\rho}g_{\mu]\nu} - i\frac{H}{F}\varepsilon_{E\beta\delta\nu[\mu}{T_{\rho]}}^{\beta\delta} 
    &+& Q_{\rho\mu\nu} -\hat{Q}_{\mu}g_{\nu\rho} + Q_{[\mu}g_{\rho]\nu} - i\frac{H}{F}\varepsilon_{E\delta\alpha\mu\nu}{Q^{\delta\alpha}}_{\rho} \notag \\
    &=& \frac{2F_{f}g_{\nu[\mu}\partial_{\rho]}f - i(H_{f}-Y_{f})\varepsilon_{E\rho\mu\nu\alpha}\partial^{\alpha}f}{F}
    , \label{eq:Cartanequation}
\end{eqnarray}
where $\mathcal{T}_{\mu\nu}$ is the energy-momentum tensor defined by
\begin{align}
    \mathcal{T}_{\mu\nu} = -\frac{1}{2}\partial_{\mu}f\partial_{\nu}f + \frac{1}{2}f^2\partial_{\mu}\theta\partial_{\nu}\theta + \frac{1}{2}g_{\mu\nu}\Bigl(\frac{1}{2}\partial_{\rho}f\partial^{\rho}f - \frac{1}{2}f^2\partial_{\rho}\theta\partial^{\rho}\theta + V\Bigr).
\end{align}
In Eq.~\eqref{eq:Einstein_equation}, we define the derivative with respect to $f$ as $O_{f} \equiv dO / df$.
From Eq.~\eqref{eq:Cartanequation} and the general form of its solution,
\begin{align}
 {T^{\mu}}_{\nu\rho} = 2t_1\delta^{\mu}_{[\nu}\partial_{\rho]}f + t_2{\varepsilon_{E\delta\nu\rho}}^{\mu}\partial^{\delta}f,\\
 Q_{\mu\nu\rho} = q_1g_{\nu\rho}\partial_{\mu}f + 2q_2g_{\mu\lbrace\nu}\partial_{\rho\rbrace}f,
\end{align}
we can express the torsion and non-metricity in terms of the radial component of~$\Phi$,
\begin{align}
    2t_1 + q_1 &= -\frac{\lbrace FF_{f} + H(H_{f} - Y_{f}) \rbrace}{F^2 + H^2},\\
    t_2 &= i\frac{\lbrace HF_{f} - F(H_{f} - Y_{f}) \rbrace}{F^2 + H^2},\\
  q_2 &= 0.
\end{align}

Here, we comment on the relation between MAG and the Einstein--Cartan theory \cite{Cartan:1923zea, Kibble:1961ba, Sciama:1964wt, Hehl:1976kj} in terms of projective invariance.
The action in Eq.~\eqref{eq:euclidean_action} remains invariant under the transformation of the affine connection, ${\Gamma^{\mu}}_{\nu\rho} \rightarrow {\Gamma^{\mu}}_{\nu\rho} + \delta^{\mu}_{\nu}\mathcal{P}{\partial}_\rho f$, where $\mathcal{P}$ is an arbitrary scalar.
This property is known as projective invariance~\cite{Hehl:1994ue}.
Indeed, the quantities $2t_1 + q_1$, $t_2$, and $q_2$ are invariant under this transformation, since the variations of the torsion and the non-metricity are given by $\delta{T^{\mu}}_{\nu\rho} = 2\delta^{\mu}_{[\nu}\mathcal{P}{\partial}_{\rho]} f$ and $\delta Q_{\mu\nu\rho} = -2\mathcal{P}({\partial}_{\mu} f) g_{\nu\rho}$, respectively.
Therefore, we can set either $q_1=0$ or $t_1=0$ by using the transformation rules $t_1\rightarrow t_1 + \mathcal{P}$ and $q_1 \rightarrow q_1 -2\mathcal{P}$.
Consequently, the subsequent results can equivalently be obtained in the Einstein--Cartan theory.

Since only these projectively invariant quantities appear in the EoMs and action, we denote them by $A$ and $B$,
\begin{align}
    A &:= 2t_1 + q_1, \\
    B &:= -it_2.
\end{align}
As a result, the curvature scalar, the Holst term, and the Nieh--Yan term are represented as
\begin{align}
    R &=  R^{\diamond} + 3\nabla^{\diamond}_{\mu}(A\partial^{\mu}f) - \frac{3}{2}(A^2 - B^2)\dot{f}^2 ,\\
    i\varepsilon_E R &= -6\nabla^{\diamond}_{\mu}(B\partial^{\mu}f) +6AB\dot{f}^2 , \\
    i\nabla^{\diamond}_{\mu}\hat{T}^{\mu} &= -6\nabla^{\diamond}_{\mu}(B\partial^{\mu}f).
\end{align}
In the spherically symmetric spacetime in \eqref{eq:metric},
the EoM for $f$ is
\begin{align}
    \ddot{f} + 3\frac{\dot{a}}{a}\dot{f} + \frac{F_{f}}{2}R + i\frac{H_f}{4}\epsilon_E R - i\frac{Y_{f}}{4}\nabla^{\diamond}_{\mu}\hat{T}^{\mu} + f\dot{\theta}^2 - \lambda f(f^2 - f_a^2) = 0, \label{eq:radial_EoM}
\end{align}
where the dot denotes the derivative with respect to $r$, so that $\dot{a} \equiv {da} / {dr}$.
The $00$- and $IJ$-components of the Einstein equation~\eqref{eq:Einstein_equation}, together with the Cartan equation~\eqref{eq:Cartanequation} and the EoM of the radial field~\eqref{eq:radial_EoM}, lead to a set of coupled second-order ordinary differential equations for the scale factor $a$ and the radial component $f$.
To simplify the analysis, we introduce the rescaled variable $\mathfrak{a}$, and obtain the EoMs and the action,
\begin{align}
    \ddot{\mathfrak{a}} &= -2\frac{\dot{\mathfrak{a}}^2}{\mathfrak{a}} + \frac{{F}_{f}}{2{F}}\dot{\mathfrak{a}}\dot{f}+\frac{2F}{M^2_p\mathfrak{a}}-\frac{\mathfrak{a}V}{F},
    \label{eq:actual_EoM1}\\
        \left(\ddot{f}+3\frac{\dot{\mathfrak{a}}}{\mathfrak{a}}\dot{f}\right)(1+G)\nonumber
        &=\frac{6(M^2_p\dot{\mathfrak{a}}^2-F)}{M^2_p\mathfrak{a}^2f}\left({F}-\frac{{F}_{f}f}{2}\right)+\frac{\dot{f}^2}{2}\left\lbrace-G_{f}+\left(\frac{3{F}_{f}}{F}-\frac{2}{f}\right)(1+G)\right\rbrace\nonumber\\
        &-\left( \frac{3{F}_{f}}{F}-\frac{2}{f} \right)V+V_{f},
        \label{eq:actual_EoM2}
\end{align}
\begin{align}
    \begin{aligned}
        S^E &= S^E_{\text{bulk}}+S^E_{\text{boundary}},\\
        S^E_{\text{bulk}}
        &=\int \sqrt{g}\,d^4x\left\{
        3F\frac{\ddot{\mathfrak a}}{\mathfrak a}
        -\frac{3}{2}F_f\frac{\dot{\mathfrak a}}{\mathfrak a}\dot f
        +(1+G)\dot f^2\right\},\\
        S^E_{\text{boundary}}
        &=-\frac{3}{2}\int d\Omega_3\,
        \left[\frac{M_p^3 \mathfrak{a}^3 Y B \dot f}{F^{3/2}}\right]_{0}^{\infty}
    \end{aligned}
    \label{eq:action1}
\end{align}
where
\begin{align}
    \mathfrak{a}&\equiv a\sqrt{{F}/M^2_p} = a \left(1 + \frac{\xi (f^2 - f_a^2)}{M_p^2}\right)^{\frac{1}{2}},\\
    G&\equiv\frac{3({F}_{f}{H}-{F}({H}_{f}-{Y}_{f}))^2}{2{F}({F}^2+{H}^2)}.\label{eq:G}
\end{align}
At the throat, the boundary condition $\dot f(0)=0$ is imposed. 
In the asymptotically flat region, the scale factor behaves as $a\to r$, while the radial field approaches and freezes at its vacuum value, $f\to f_a,~\dot{f}(\infty)=0$.
Thus, the boundary term $S^E_{\text{boundary}}$ in Eq.~\eqref{eq:action1} vanishes, so that the action reduces to the bulk contribution, $S^E=S^E_{\text{bulk}}$.
The action and the dynamics of the wormhole are then determined by the non-minimal coupling parameter $\xi$ and the factor $G$~\eqref{eq:G}, while the initial value $\mathfrak{a}(0)$ is determined by
\begin{align}
    3{F(0)}^2 \mathfrak{a}(0)^4=\frac{q{F(0)}^3}{\lambda^2M^4_pf(0)^2}+\frac{\lambda M^2_p\mathfrak{a}(0)^6(f(0)^2-f_a^2)^2}{4},
    \label{eq:actual_initial}
\end{align}
through the boundary conditions of the axionic wormhole.
The parameter $q$ in \eqref{eq:actual_initial} is defined by $q \equiv {\lambda^2 n^2} / {8 \pi^4}$.

\section{Numerical Analysis}
\label{sec:NA}

In this section, we numerically investigate the axionic wormhole solution by solving Eqs.~\eqref{eq:actual_EoM1} and \eqref{eq:actual_EoM2}.
They are second-order differential equations for the scale factor $a$ and the radial component $f$ of $\Phi$ with respect to the radial coordinate $r$.
We impose the following boundary conditions:
\begin{align}
\label{eq:boundary}
    \dot{a}(0)=0,\quad \dot{f}(0)=0, \quad f(\infty) = f_a.
\end{align}
Using Eq.~\eqref{eq:actual_initial} to determine $a(0)$ from $f(0)$, we tune these initial values to find a finite solution that satisfies the boundary conditions and is asymptotically flat.
In the following calculation, we adopt the dimensionless variables~\cite{Abbott:1989jw, Hamaguchi:2021mmt,Cheong:2022ikv},
\begin{align}
    \rho &\equiv \sqrt{3\lambda}M_p\, r, \quad \mathcal{A} \equiv \sqrt{3\lambda}M_p \,a,\quad \mathcal{F} \equiv \frac{f}{\sqrt{3}M_p}.
\end{align}
We also fix the parameters to $n =1,~\lambda = 0.1$, and $f_a = 10^{15}$~GeV.

As discussed in the previous section, MAG introduces five independent parameters (the non-minimal couplings and the dimensionless parameters $\gamma_H$, $\gamma_\eta$).
To clarify how these couplings affect the wormhole solution and the resulting axion quality problem, we systematically analyze their effects.
First, we examine cases where only a single non-minimal coupling is present, so that its characteristic effect can be isolated.
Subsequently, we consider the simultaneous presence of two couplings to investigate how their interplay modifies the wormhole dynamics.
In our analysis, we present the axionic wormhole solutions for specific parameter values as benchmarks.
For a more detailed discussion of the wormhole profiles, see Appendix~\ref{sec:F0A0}, where we examine how the initial values of $\mathcal F$ and $\mathcal A$ depend on each coupling parameter.
Also, as in previous works~\cite{Hamaguchi:2021mmt, Cheong:2022ikv,  Cheong:2024kum}, we check whether the parameter regions that alleviate the axion quality problem are compatible with inflation.
For this inflationary analysis, we adopt the method used in Refs.~\cite{Langvik:2020nrs, Shaposhnikov:2020gts} and incorporate observational constraints from the Atacama Cosmology Telescope~(ACT) results \cite{BICEP:2021xfz, AtacamaCosmologyTelescope:2025blo, AtacamaCosmologyTelescope:2025nti}. 
For comparison with these constraints, we use the benchmark conditions
\begin{align}
    N_e=80,\quad 0.97<n_s<0.98,\quad r<0.038, \label{eq:InflationConstraint}
\end{align}
where $N_e$, $n_s$, and $r$ denote the number of e-folds, the spectral index, and the tensor-to-scalar ratio, respectively.
The details of the inflationary analysis are summarized in Appendix~\ref{sec:inflation}.

\subsection{Single-coupling cases}
\label{sec:single-coupling}

In this subsection, we discuss three simple cases where the field is non-minimally coupled only to the curvature scalar, the Nieh--Yan term, or the Holst term.
We find that the axion quality problem can be alleviated in all three cases: the threshold is characterized by $\xi$ for the curvature scalar, $\chi_{\eta} \equiv {\xi_{\eta}} / {\gamma_{\eta}}$ for the Nieh--Yan term, and $\chi_H \equiv {\xi_H} / {\gamma_H}$ (at fixed $\gamma_H$) for the Holst term.
The threshold values in the single-coupling cases are summarized in Table~\ref{tab:thresholds}.

\begin{table}[t]
    \centering
    \begin{tabular}{c c}
        \hline
        Parameter & Threshold value \\
        \hline
        $\xi$ & $\xi^{\rm th} = 9.4\times10^3$ \\
        $\chi_{\eta} \equiv {\xi_{\eta}} / {\gamma_{\eta}}$ & $\chi_\eta^{\rm th} = 2.5\times 10^3$ \\
        $\chi_H \equiv {\xi_H} / {\gamma_H}$ with $\gamma_H=10$ & $\chi_H^{\rm th} = 2.7\times 10^3$ \\
        \hline
    \end{tabular}
    \caption
    {Threshold values in the single-coupling cases with $n =1,~\lambda = 0.1$, and $f_a = 10^{15}$~GeV.
    }
    \label{tab:thresholds}
\end{table}

\subsubsection{Non-minimal coupling to the curvature scalar}
\label{sec:Non_minimal}

First, we consider a non-minimal coupling to the curvature scalar, which is the limit $\gamma_{H}, \gamma_{\eta}\rightarrow\infty$.
This corresponds to the Palatini formalism because we can set $A=-F_{f}/F$, $B= G = t_1 = 0$ using projective invariance.
The axionic wormhole solutions in the Palatini formalism were investigated in Ref.~\cite{Cheong:2022ikv}, and this case can alleviate the axion quality problem for $\xi \gtrsim 9.4 \times 10^{3}$.

\subsubsection{Non-minimal coupling to the Nieh--Yan term}
\label{sec:Nieh-Yan}

\begin{figure}[t]
  \centering
  \includegraphics[width=0.48\textwidth]{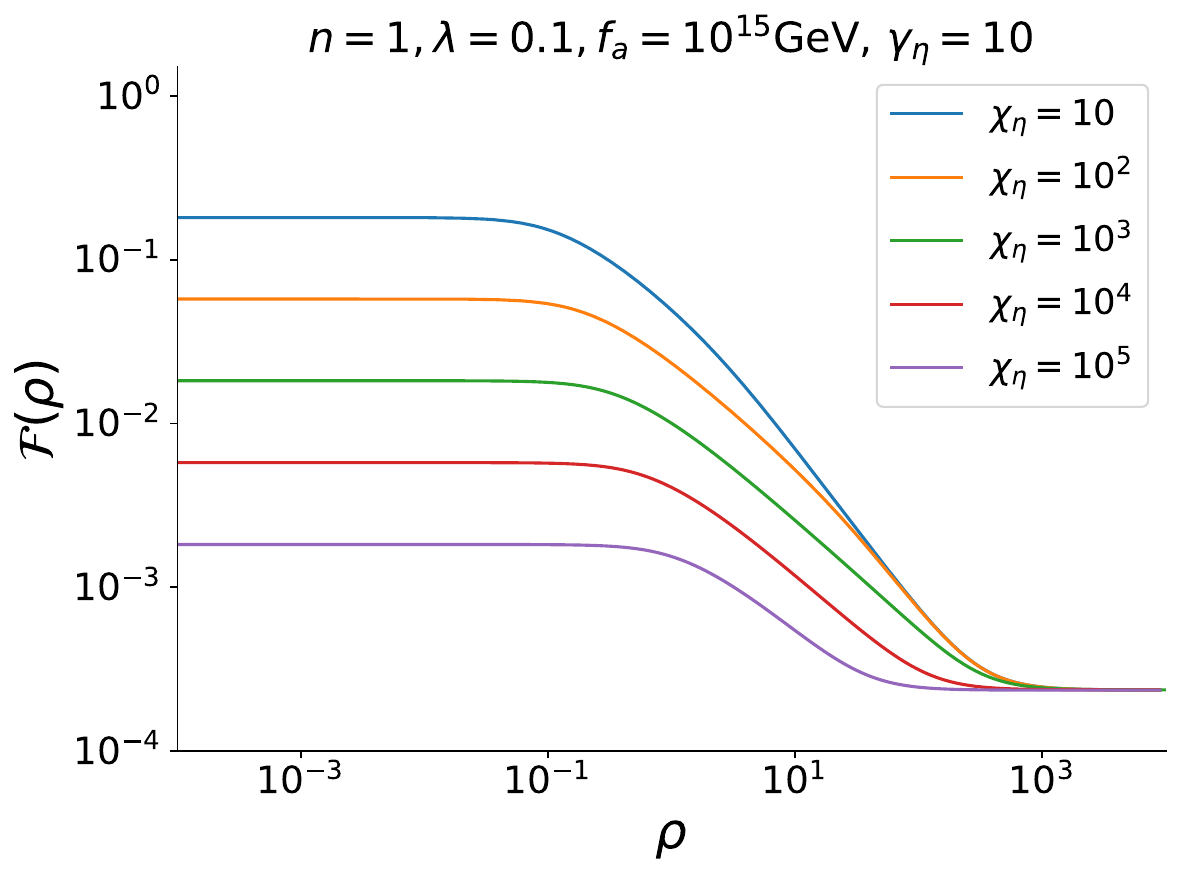}
  \hfill
  \includegraphics[width=0.48\textwidth]{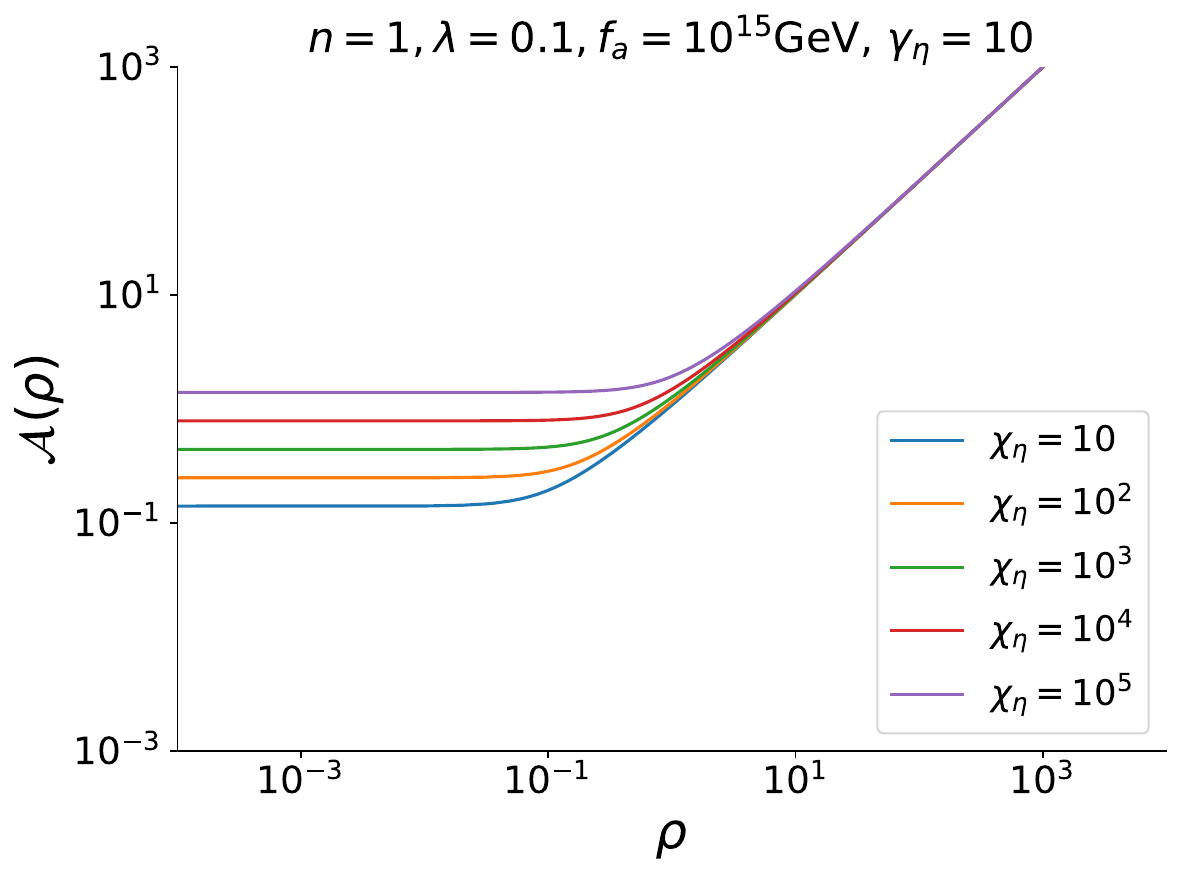}
  \caption{$\mathcal{F}(\rho)$ and $\mathcal{A}(\rho)$ for several values of $\chi_\eta = 10^1,~10^2,~10^3,~10^4$, and $10^5$ with $\gamma_\eta = 10$.}
  \label{fig:FA_Nieh}
\end{figure}

Next, we consider a non-minimal coupling to the Nieh--Yan term, which is the limit $\xi\rightarrow 0,~\gamma_{H}\rightarrow\infty$.
In this case, the factor $G$ in Eq.~\eqref{eq:G} takes the form,
\begin{align}
\label{eq:G_Nieh}
    G = 18 \chi_{\eta}^2{\mathcal F}^2, 
\end{align}
where $\chi_{\eta}$ is defined by $\chi_{\eta} \equiv \xi_{\eta}/\gamma_{\eta}$.
Therefore, the action and the dynamics of the wormhole depend only on $\chi_{\eta}$. 
In Fig.~\ref{fig:FA_Nieh}, we show the axionic wormhole solutions of $\mathcal{F}(\rho)$ and $\mathcal{A}(\rho)$ for several values of $\chi_\eta$ with $\gamma_{\eta} = 10$.
As expected from the above discussion, the behavior of these solutions depends on $\chi_\eta$. 
Specifically, as $\chi_\eta$ increases, the wormhole throat $\mathcal{A}(0)$ increases while the initial radial field value $\mathcal{F}(0)$ decreases.

\begin{figure}[tbp]
  \centering
  \includegraphics[width=0.7\textwidth]{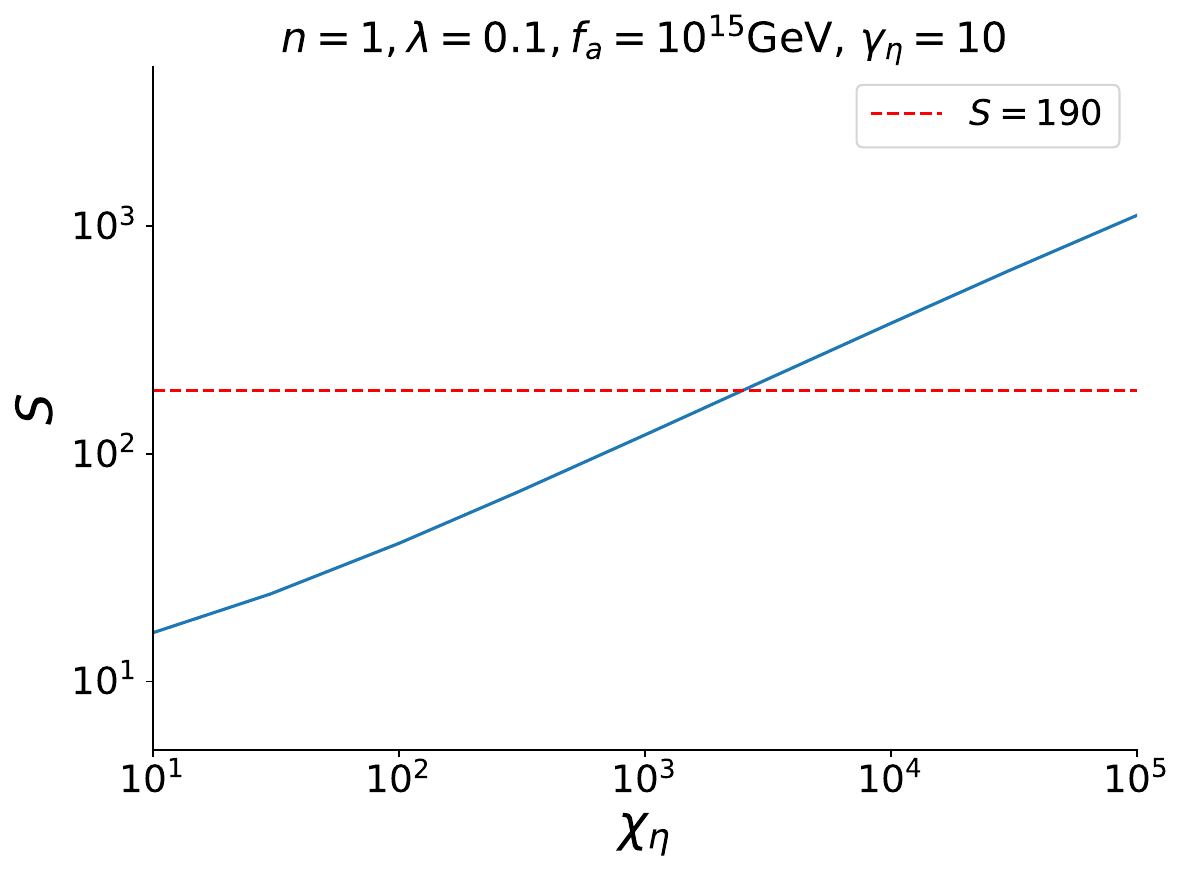}
\caption{
The value of the action $S$ as a function of $\chi_\eta$.
}
\label{fig:S_Nieh}
\end{figure}

Fig.~\ref{fig:S_Nieh} shows the value of the axionic wormhole action $S$ as a function of $\chi_\eta$.
The blue line corresponds to the results for $\gamma_\eta = 10$.
The horizontal red dashed line depicts the threshold required to solve the axion quality problem, $S = 190$~\cite{Kallosh:1995hi, Alvey:2020nyh}.
As demonstrated in Fig.~\ref{fig:S_Nieh}, the value of the action $S$ increases with $\chi_\eta$.
To solve the axion quality problem with the non-minimal coupling only to the Nieh--Yan term, we require $\chi_\eta \gtrsim 2.5 \times 10^3$.

\subsubsection{Non-minimal coupling to the Holst term}
\label{sec:Holst}

\begin{figure}[t]
  \centering
  \includegraphics[clip, width=0.48\textwidth]{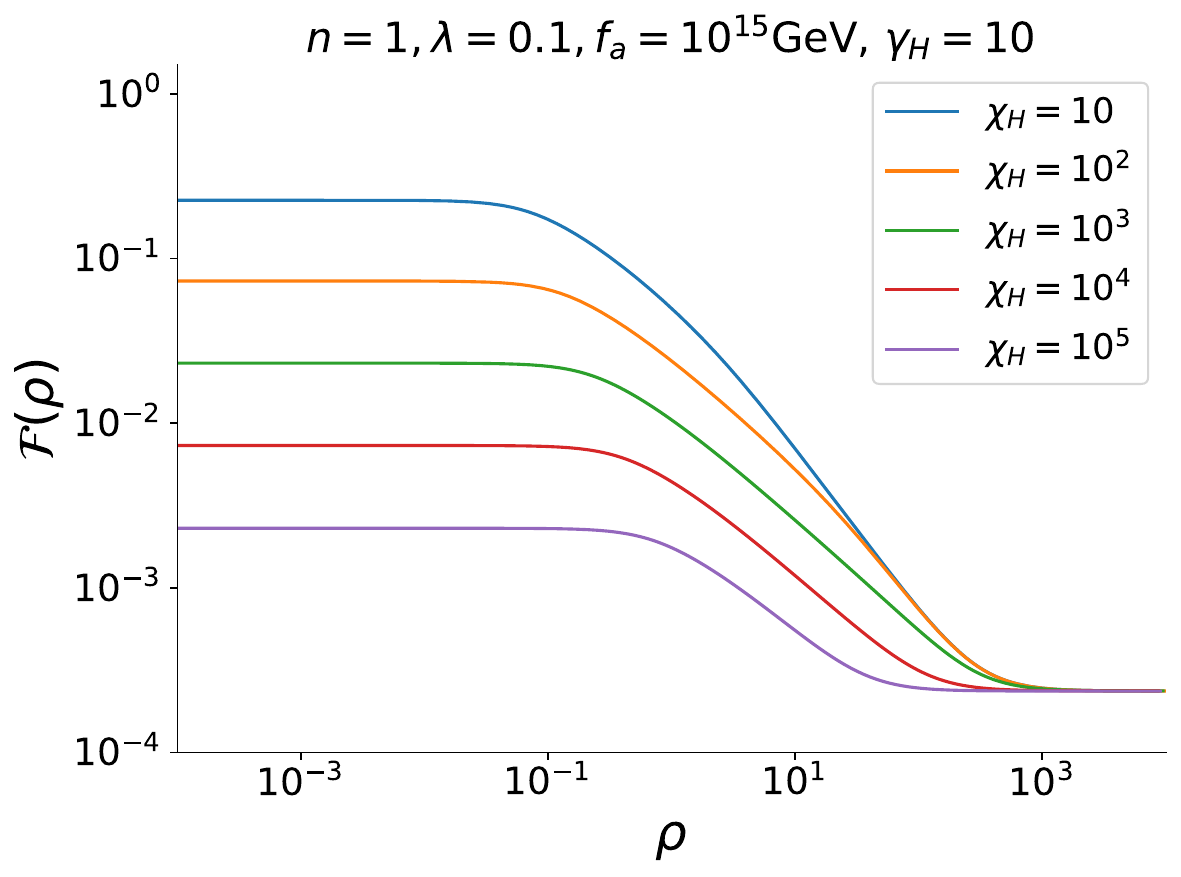}
  \hfill
  \includegraphics[clip, width=0.48\textwidth]{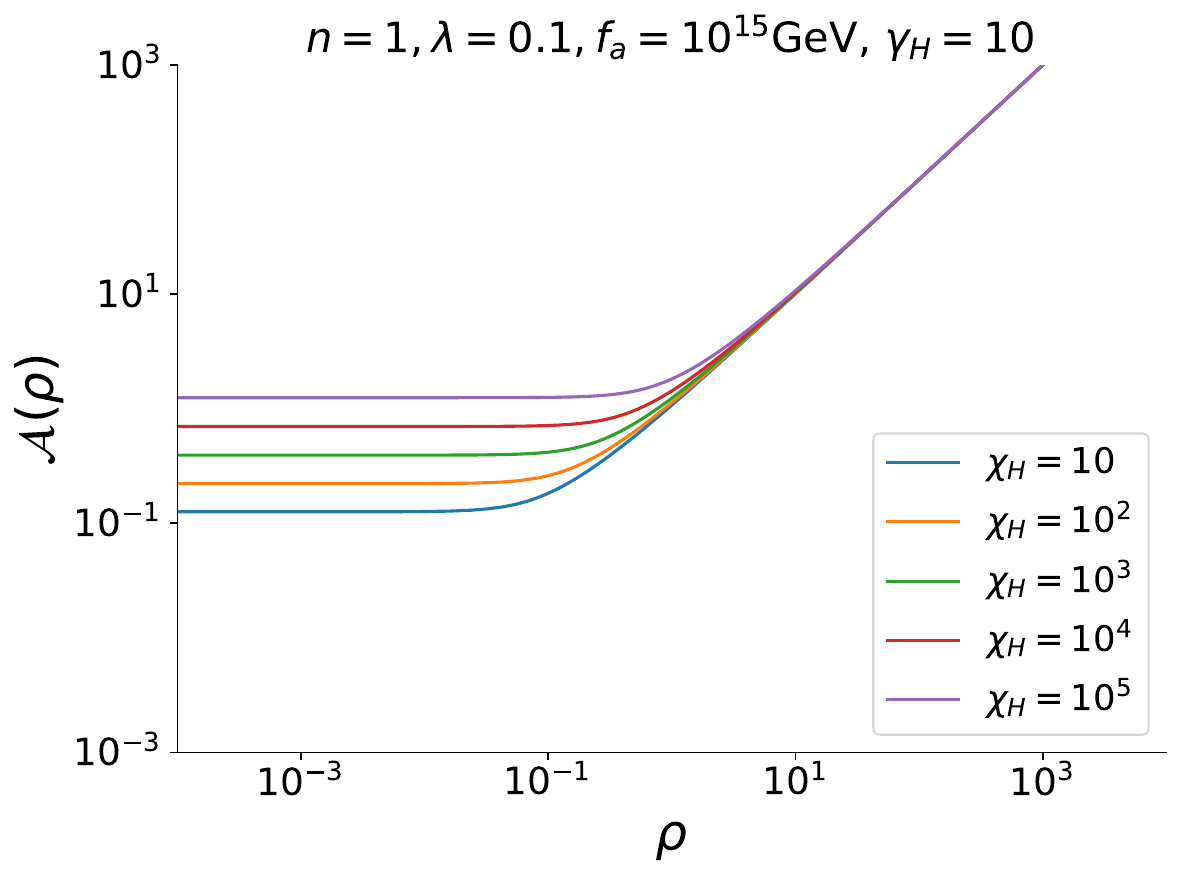}
  \caption{$\mathcal{F}(\rho)$ and $\mathcal{A}(\rho)$ for several values of $\chi_H = 10^1,~10^2,~10^3,~10^4$, and $10^5$ with $\gamma_H = 10$.}
  \label{fig:FA_Holst}
\end{figure}

\begin{figure}[t]
  \centering
  \includegraphics[clip, width=0.48\textwidth]{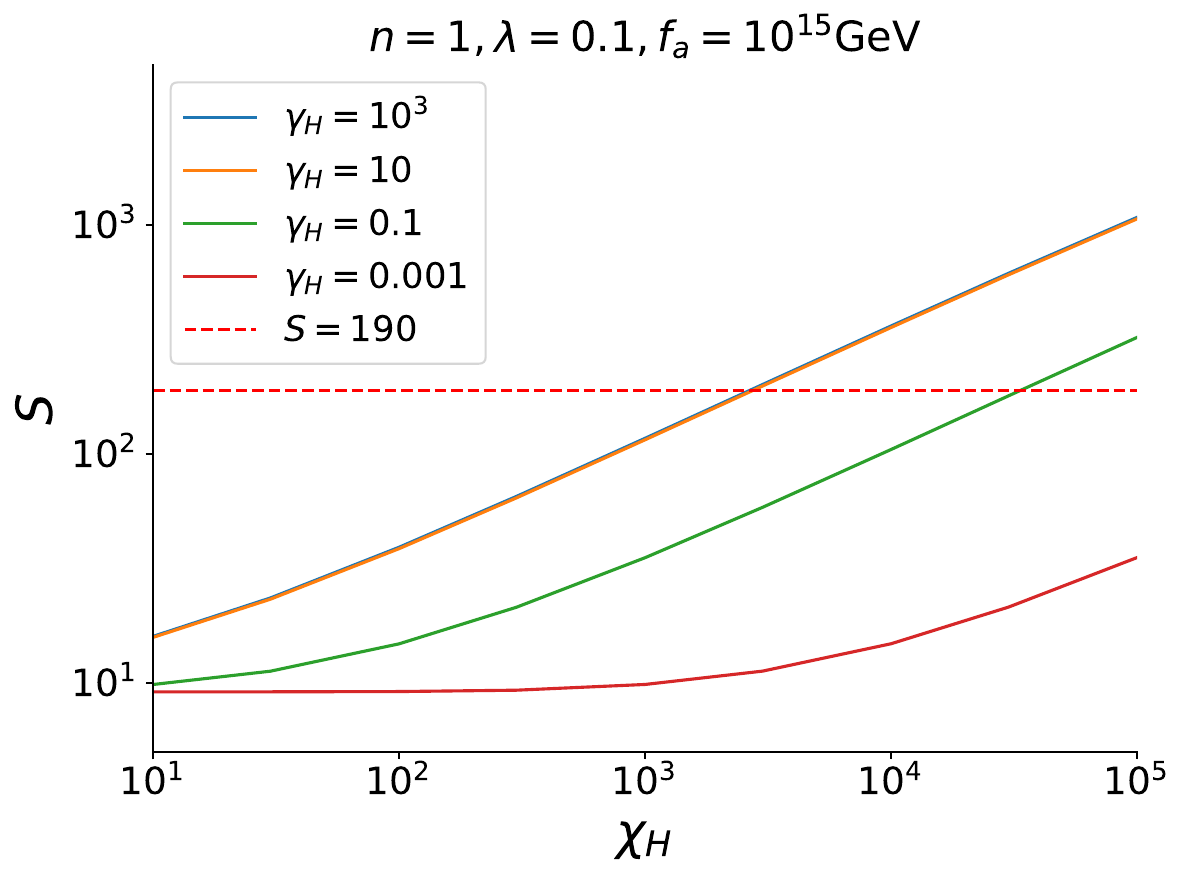}
  \hfill
  \includegraphics[clip, width=0.48\textwidth]{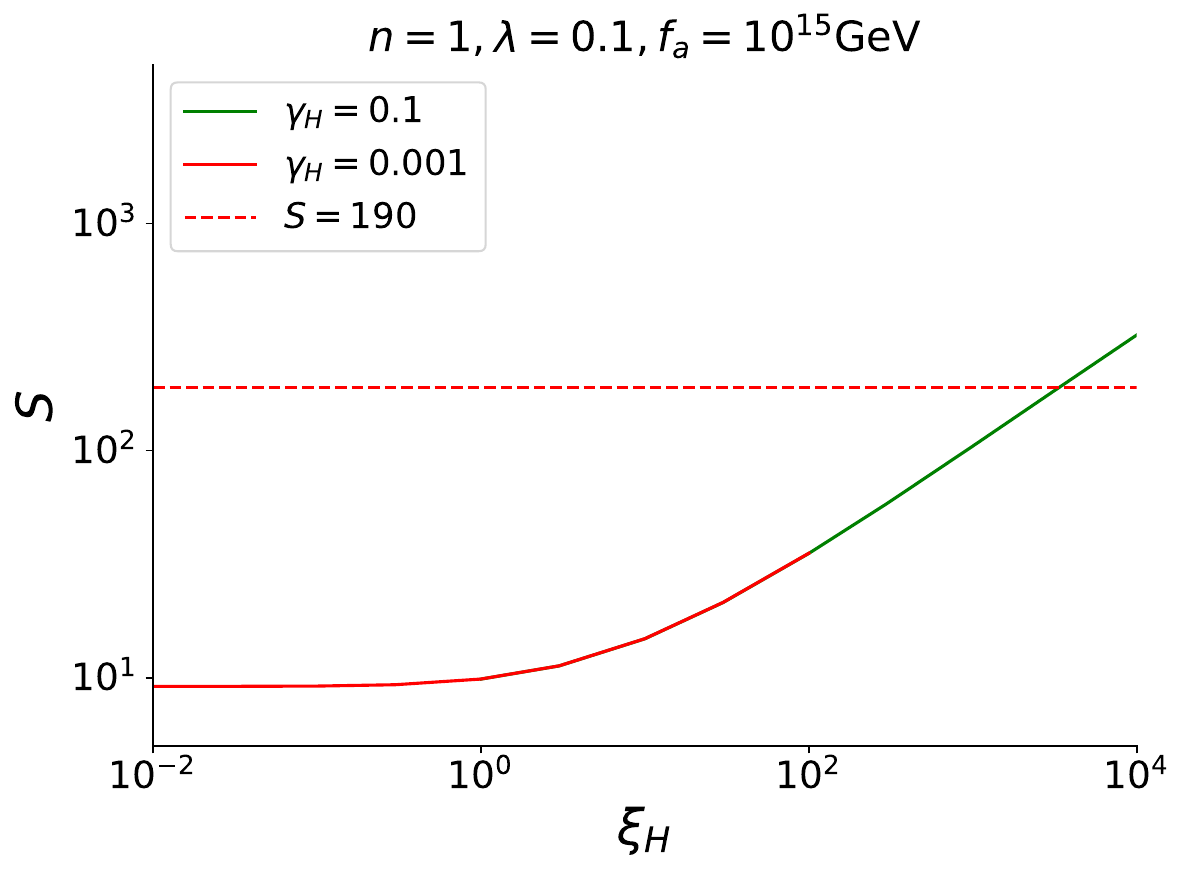}
  \caption{The value of the action $S$ as a function of $\chi_H$ (left) and $\xi_H$ (right).}
  \label{fig:S_Holst}
\end{figure}

Finally, we consider a non-minimal coupling to the Holst term, which is the limit $\xi\rightarrow 0,~\gamma_{\eta}\rightarrow\infty$.
In this case, the factor $G$ in Eq.~\eqref{eq:G} takes the form,
\begin{align}
    G = \frac{18 \xi_{H}^2 {\mathcal F}^2}{\gamma_H^2+(1 + 3\xi_H(\mathcal{F}^2-\mathcal{F}^2_a))^2},
\end{align}
where $\mathcal{F}_a \equiv f_a / \left(\sqrt{3} M_p\right)$.
In this case, the behavior of the wormhole solutions differs between the large $\gamma_H$ and small $\gamma_H$ regimes.
In the large $\gamma_H$ case, the factor $G$ is approximately given by
\begin{align}
\label{eq:G_Holst_large}
    G \sim 18 \chi_{H}^2 {\mathcal F}^2,
\end{align}
where $\chi_{H}$ is defined by $\chi_{H} \equiv \xi_{H}/\gamma_{H}$.
In this case, the dynamics of the wormhole depends only on $\chi_{H}$; hence, a large value of $\chi_H$ is necessary for solving the axion quality problem.
On the other hand, in the small $\gamma_H$ case, the factor $G$ is approximately
\begin{align}
\label{eq:G_Holst_small}
    G \sim \frac{18 \xi_{H}^2 {\mathcal F}^2}{(1 + 3\xi_H(\mathcal{F}^2-\mathcal{F}^2_a))^2}.
\end{align}
Thus, the wormhole action depends only on $\xi_H$, and a large value of $\xi_H$ is necessary for solving the axion quality problem in the small $\gamma_H$ case.

Fig.~\ref{fig:FA_Holst} shows the axionic wormhole solutions of $\mathcal{F}(\rho)$ and $\mathcal{A}(\rho)$ for several values of $\chi_H$ with $\gamma_H = 10$.
As expected from the above discussion, these results for large $\gamma_H$ exhibit a behavior that mostly depends on $\chi_H$.
As $\chi_H$ increases, the wormhole throat $\mathcal{A}(0)$ increases while the initial radial field value $\mathcal{F}(0)$ decreases.
These behaviors are quite similar to those in the case where only the Nieh--Yan term is non-minimally coupled.
In contrast, for small $\gamma_H$, the behaviors of the wormhole solutions depend only on $\xi_H$, as discussed in Appendix~\ref{sec:F0A0}.

In Fig.~\ref{fig:S_Holst}, we show the value of the axionic wormhole action $S$ as a function of $\chi_H$ and $\xi_H$.
The blue, orange, green, and red lines indicate the results for $\gamma_H = 10^3,~10,~0.1,$ and $10^{-3}$, respectively.
The horizontal red dashed line depicts the threshold required to solve the axion quality problem, $S = 190$~\cite{Kallosh:1995hi, Alvey:2020nyh}.
As shown in the left panel of Fig.~\ref{fig:S_Holst}, in the large $\gamma_H$ case~($\gamma_H = 10^3,~10$), the value of the action $S$ depends only on $\chi_H$, and increases with $\chi_H$.
In the small $\gamma_H$ case~($\gamma_H = 0.1,~10^{-3}$), as shown in the right panel of Fig.~\ref{fig:S_Holst}, the value of the action $S$ depends only on $\xi_H$, and increases with $\xi_H$.
Therefore, in the case with only the Holst term, the axion quality problem can be solved for $\chi_H \gtrsim 2.7 \times 10^3$ in the large $\gamma_H$ regime.
In the small $\gamma_H$ regime, $\chi_H$ must be larger than in the large $\gamma_H$ regime to solve this problem.

Before turning to the two-coupling cases, we comment on the origin of the small difference between $\chi_H^{\rm th}$ and $\chi_\eta^{\rm th}$ in the large $\gamma_H$ regime.
Compared with Eq.~\eqref{eq:G_Nieh}, the numerators of $G$ coincide under $\chi_\eta \leftrightarrow \chi_H$, and the two cases differ only through the denominator, i.e.\ the replacement $F^2 \to F^2+H^2$ in Eq.~\eqref{eq:G}.
This denominator suppresses $G$ and thus makes $\chi_H^{\rm th}$ slightly larger than $\chi_\eta^{\rm th}$ in Table~\ref{tab:thresholds}.

\subsection{Two-coupling cases}
\label{sec:two-coupling}

In this subsection, we discuss the two-coupling scenarios: the combinations of the curvature scalar and the Nieh--Yan term, the curvature scalar and the Holst term, and the Nieh--Yan and Holst terms.
We show that in all three cases, the axion quality problem can be solved over a broader region of parameter space than in the single-coupling cases considered above.
We also explore whether the regions where the axion quality problem is resolved can overlap with those compatible with inflationary constraints.

\subsubsection{The curvature scalar and the Nieh--Yan term}
\label{sec:curvatureNieh}

\begin{figure}[t]
  \centering
  \includegraphics[clip, width=0.48\textwidth]{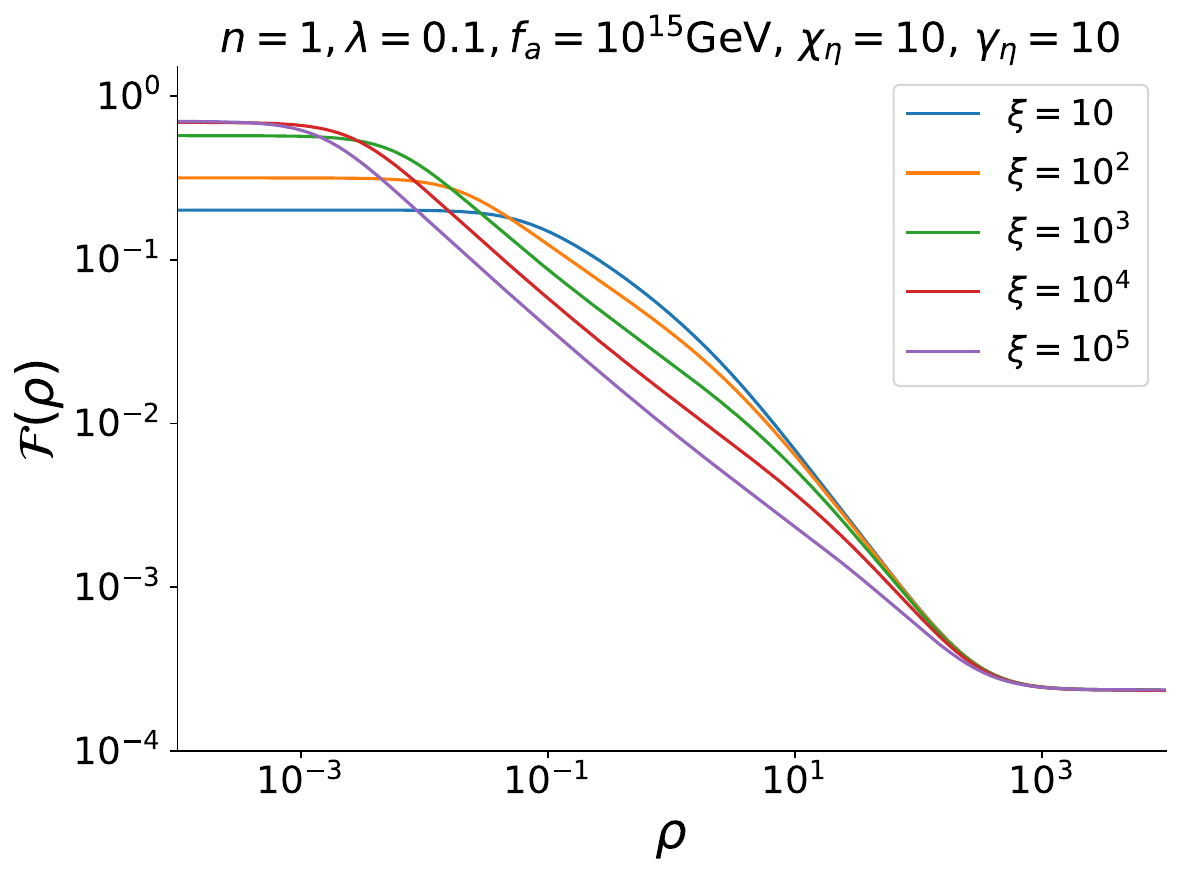}
  \hfill
  \includegraphics[clip, width=0.48\textwidth]{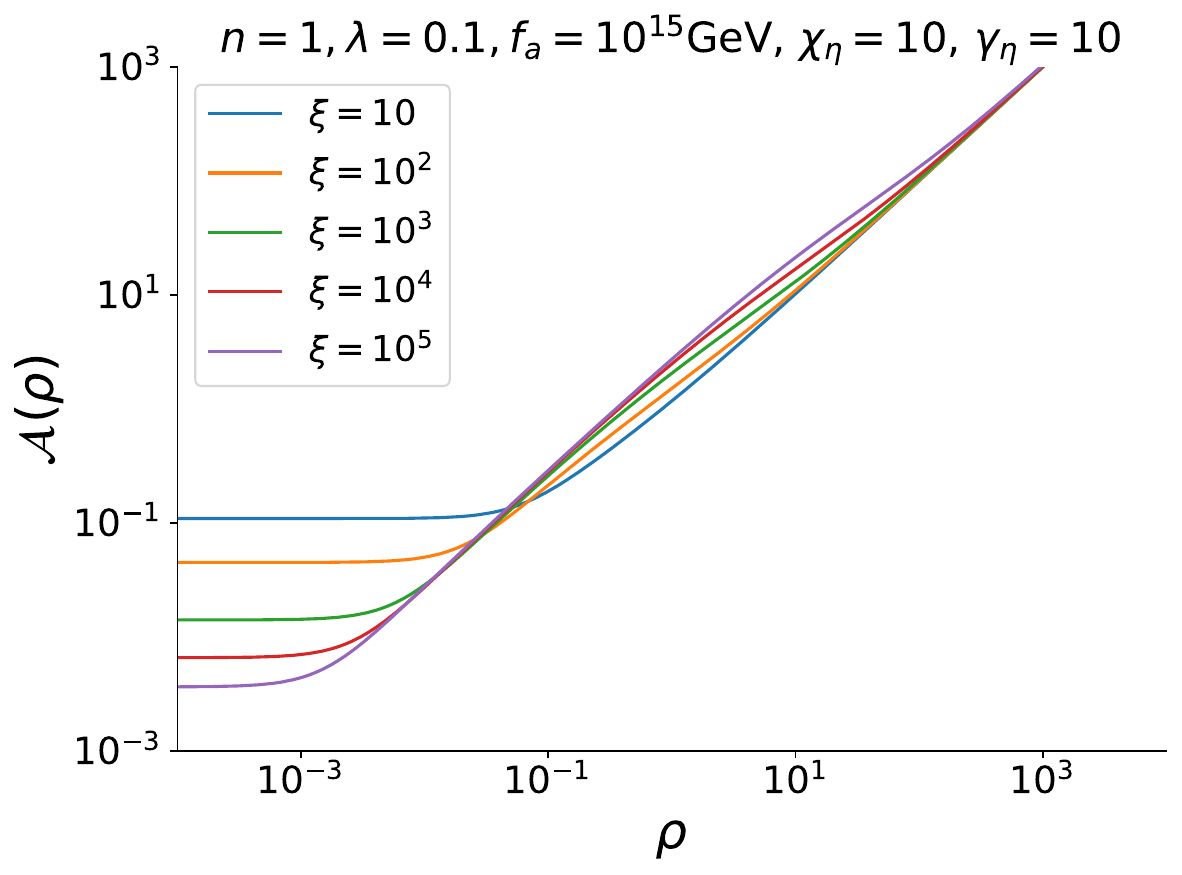}
  \caption{$\mathcal{F}(\rho)$ and $\mathcal{A}(\rho)$ for several values of $\xi = 10^1,~10^2,~10^3,~10^4$, and $10^5$ with $\chi_\eta = 10$ and $\gamma_\eta = 10$.}
  \label{fig:FA_nonminimal_Nieh}
\end{figure}

\begin{figure}[t]
  \centering
  \includegraphics[width=0.7\textwidth]{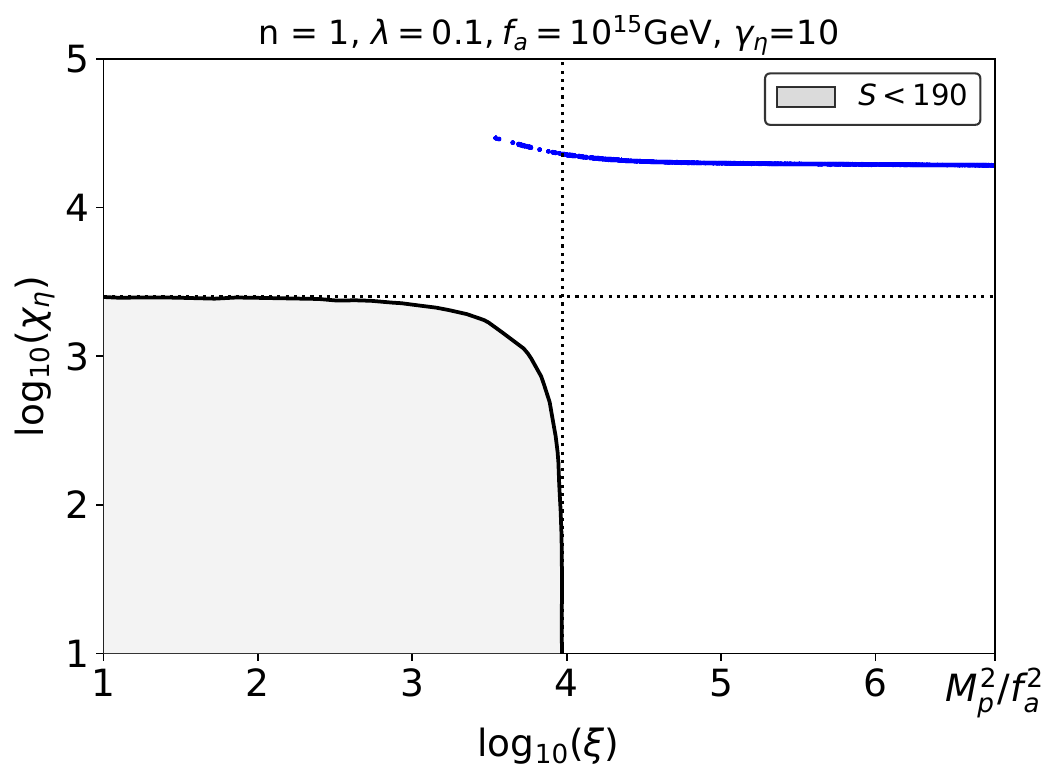}
\caption{
The parameter region in the $(\xi,\chi_\eta)$ plane for $\gamma_\eta=10$.
The gray shaded region corresponds to $S < 190$, where the axion quality problem remains unsolved.
The black dotted lines indicate the thresholds required to solve the axion quality problem in each single-coupling limit.
The blue line denotes the inflation-compatible trajectory satisfying $N_e=80$, $0.97<n_s<0.98$, and $r<0.038$.
}
\label{fig:WormInfNieh}
\end{figure}

First, we discuss non-minimal couplings to the curvature scalar and the Nieh--Yan term.
This case is obtained in the limit $\gamma_{H}\rightarrow\infty$.
The factor $G$ in Eq.~\eqref{eq:G} takes the form,
\begin{align}
\label{eq:G_nonminimal_Nieh}
    G = \frac{18 \chi_{\eta}^2 {\mathcal F}^2}{(1+ 3\xi(\mathcal{F}^2-\mathcal{F}^2_a))} .
\end{align}
In the large $\xi$ case, the factor $G$ is approximately given by $G \propto {\chi^2_{\eta}} / {\xi}$.
If the ratio ${\chi^2_{\eta}} / {\xi}$ is small, the wormhole dynamics is expected to approach that in the Palatini formalism.
On the other hand, a small $\xi$ leads to the behavior associated with the non-minimal coupling only to the Nieh--Yan term discussed in Sec.~\ref{sec:Nieh-Yan}.

Fig.~\ref{fig:FA_nonminimal_Nieh} shows the axionic wormhole solutions of $\mathcal{F}(\rho)$ and $\mathcal{A}(\rho)$ for several values of $\xi$ with $\chi_\eta = 10$.
As anticipated from the discussion above, as $\xi$ increases, the wormhole throat $\mathcal{A}(0)$ decreases while the initial radial field value $\mathcal{F}(0)$ increases and converges to an almost universal value.
This behavior in the large $\xi$ regime ($10^2 \le \xi \le 10^5$) is consistent with that obtained in the Palatini formalism~\cite{Cheong:2022ikv}.

In Fig.~\ref{fig:WormInfNieh}, we show the region in the $(\xi,\chi_\eta)$ plane in which the axion quality problem remains unsolved, together with the parameters satisfying the inflationary requirements in Eq.~\eqref{eq:InflationConstraint}, for fixed $\gamma_\eta = 10$.
We find that the $S<190$ region is approximately rectangular, but its upper-right corner is rounded inward. 
Consequently, the axion quality problem can be resolved even when both $\xi$ and $\chi_\eta$ are below their threshold values $\xi^{\rm th}$ and $\chi_{\eta}^{\rm th}$.
The intersections of the region's boundary with the two axes are consistent with the corresponding single-coupling limits.
Furthermore, successful inflation can be simultaneously realized only along the blue line $\chi_\eta \sim 2 \times 10^4$ within the range $10^4 \lesssim \xi \lesssim M_p^2 / f_a^2$.

\begin{figure}[t]
  \centering
  \includegraphics[clip, width=0.48\textwidth]{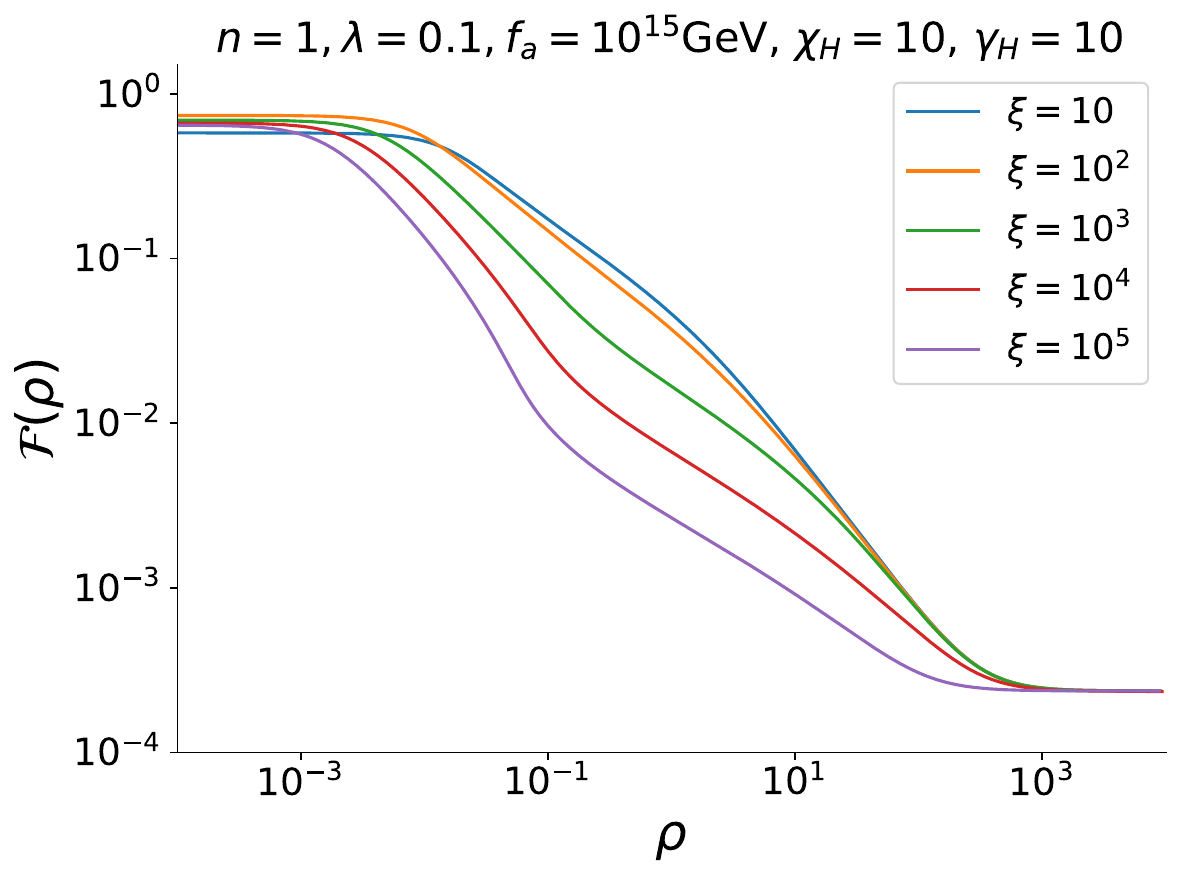}
  \hfill
  \includegraphics[clip, width=0.48\textwidth]{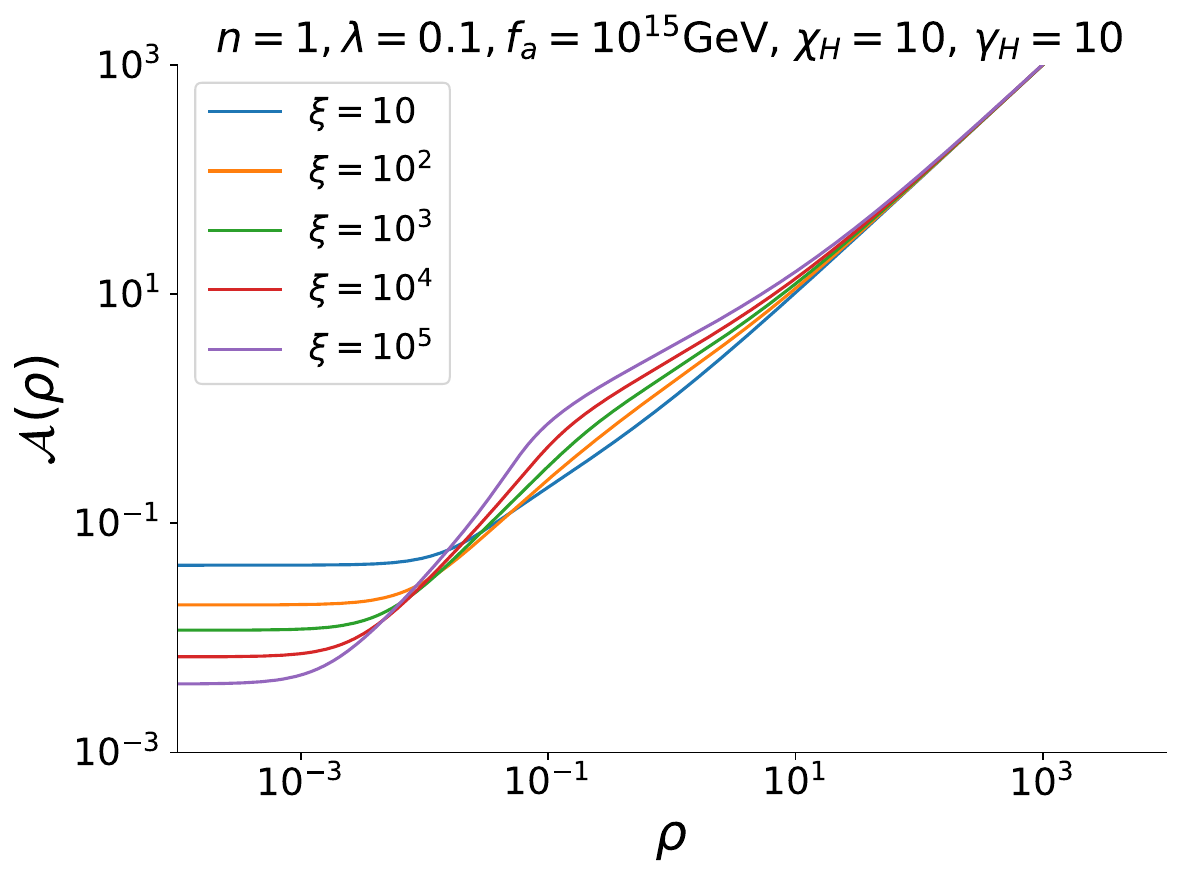}
  \caption{$\mathcal{F}(\rho)$ and $\mathcal{A}(\rho)$ for several values of $\xi = 10^1,~10^2,~10^3,~10^4$, and $10^5$ with $\chi_H = 10$ and $\gamma_H = 10$.}
  \label{fig:FA_nonminimal_Holst}
\end{figure}

\begin{figure}[t]
  \centering
  \includegraphics[width=0.7\textwidth]{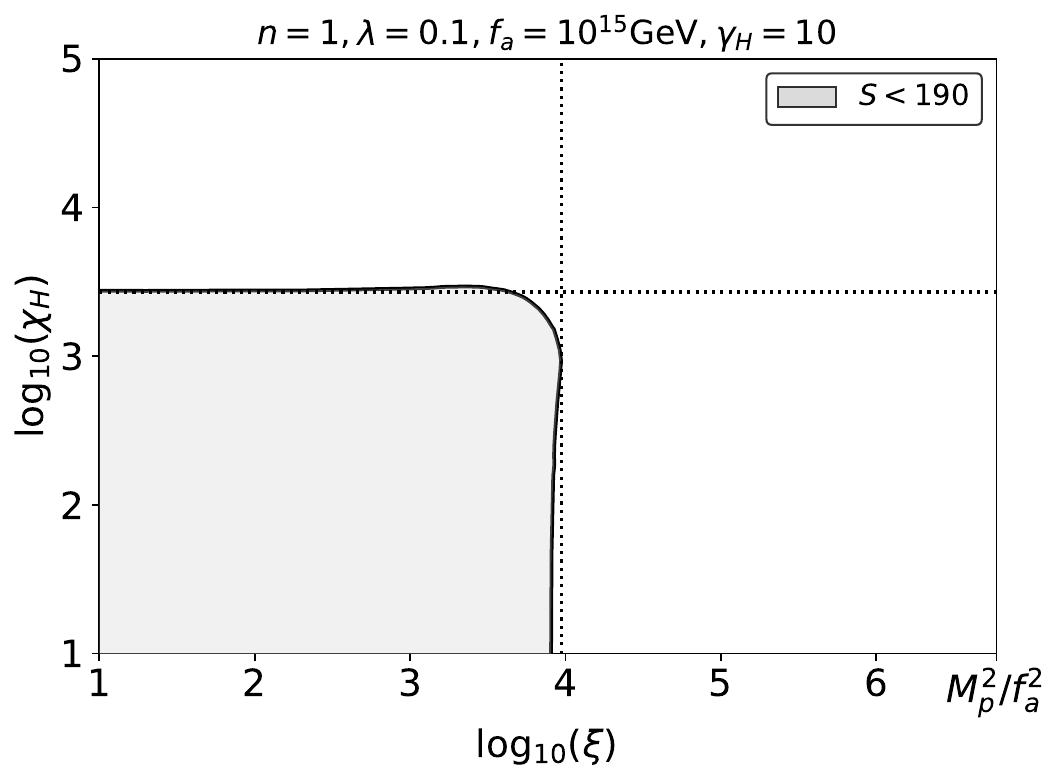}
\caption{
The parameter region in the $(\xi,\chi_H)$ plane for $\gamma_H=10$.
The gray shaded region corresponds to $S < 190$, where the axion quality problem remains unsolved.
The black dotted lines indicate the thresholds required to solve the axion quality problem in each single-coupling limit.
Inflation cannot be successfully realized in any region.
}
\label{fig:WormInfHolst}
\end{figure}

\subsubsection{The curvature scalar and the Holst term}
\label{sec:curvatureHolst}

Next, we discuss non-minimal couplings to the curvature scalar and the Holst term.
This case is obtained in the limit $\gamma_{\eta}\rightarrow\infty$.
The factor $G$ in Eq.~\eqref{eq:G} takes the form,
\begin{align}
    G = \frac{18 {\mathcal F}^2(\xi-\xi_{H})^2}{{(1 + 3\xi({\mathcal F}^2-{\mathcal F}^2_a))}(\gamma_{H}^2(1 + 3\xi({\mathcal F}^2-{\mathcal F}^2_a))^2+(1 + 3\xi_H({\mathcal F}^2-{\mathcal F}^2_a))^2)}.
    \label{eq:factorGXiHolst}
\end{align}
As shown in Eq.~\eqref{eq:factorGXiHolst}, the Barbero--Immirzi parameter $\gamma_H$ determines the contribution of the $(1+3\xi({\mathcal F}^2-{\mathcal F}^2_a))^2$ term in the denominator. 
Furthermore, it can be seen that setting $\xi = \xi_H$ yields the wormhole profile in the Palatini formalism because $G = 0$.

Fig.~\ref{fig:FA_nonminimal_Holst} shows the axionic wormhole solutions of $\mathcal{F}(\rho)$ and $\mathcal{A}(\rho)$ for several values of $\xi$ with $\chi_H=10$ and $\gamma_H=10$.
In this case, $\gamma_H$ is relatively large, and thus the $\gamma_H^2(1+3\xi(\mathcal{F}^2-\mathcal{F}_a^2))^2$ term dominates the denominator of $G$.
As a result, the dependence on the Holst non-minimal coupling parameter $\xi_H$ is suppressed.
In fact, Fig.~\ref{fig:FA_nonminimal_Holst} shows that, as $\xi$ increases, the wormhole throat $\mathcal{A}(0)$ decreases while the initial radial field value $\mathcal{F}(0)$ increases and converges to a nearly universal value.
These profiles are close to those in the Palatini formalism.
As in the case where only the Holst term is non-minimally coupled with small $\gamma_H$, the dynamics depends on $\xi_H$ rather than $\chi_H$ in the small $\gamma_H$ regime with fixed $\xi$.
In Appendix~\ref{sec:F0A0}, the behaviors of the wormhole solutions in the small $\gamma_H$ regime are discussed in detail. 

In Fig.~\ref{fig:WormInfHolst}, we show the region in the $(\xi,\chi_H)$ plane where the axion quality problem remains unsolved, for fixed $\gamma_H = 10$.
We find that compared with the $(\xi,\chi_\eta)$ case in Fig.~\ref{fig:WormInfNieh}, the boundary of the $S<190$ region is closer to a rectangle.
Along $\xi \simeq \xi_H$, the Holst contribution vanishes, $G=0$, and the model reduces to the Palatini formalism.
Near this line, the $S<190$ region slightly exceeds the threshold $\chi_H^{\rm th}$, while it nearly reaches the threshold $\xi^{\rm th}$.
Also, the intersection of the boundary with the $\xi$ axis lies below the threshold $\xi^{\rm th}$, since $\gamma_H$ is finite.
In the limit $\gamma_H \to \infty$, the contribution of the Holst term disappears, and the boundary is expected to coincide with the corresponding single-coupling limit $\xi^{\rm th}$.
As discussed in Sec.~\ref{sec:Holst}, the small $\gamma_H$ regime is governed by $\xi_H$ rather than $\chi_H$.
Since $\chi_H = \xi_H / \gamma_H$, a smaller $\gamma_H$ corresponds to a larger $\chi_H$ for a fixed $\xi_H$.
Therefore, for smaller values of $\gamma_H$, the region where $S < 190$ is expected to expand toward larger $\chi_H$.
In this case, successful inflation cannot be realized in any region.

\subsubsection{The Nieh--Yan and Holst terms}
\label{sec:NiehHolst}

Finally, we discuss non-minimal couplings to the Nieh--Yan and Holst terms.
This case is obtained in the limit $\xi \rightarrow 0$.
The factor $G$ in Eq.~\eqref{eq:G} takes the form
\begin{align}
\label{eq:G_NH}
    G = \frac{18\gamma_H^2\mathcal{F}^2(\chi_{\eta}-\chi_{H})^2}
    {\gamma_H^2+(1 + 3\xi_H(\mathcal{F}^2-\mathcal{F}^2_a))^2}.
\end{align}
Away from the cancellation line $\chi_\eta=\chi_H$, the wormhole dynamics is similar to that in the case with only the Holst term, with the relative combination $(\chi_\eta-\chi_H)$.

\begin{figure}[t]
  \centering
  \includegraphics[clip, width=0.48\textwidth]{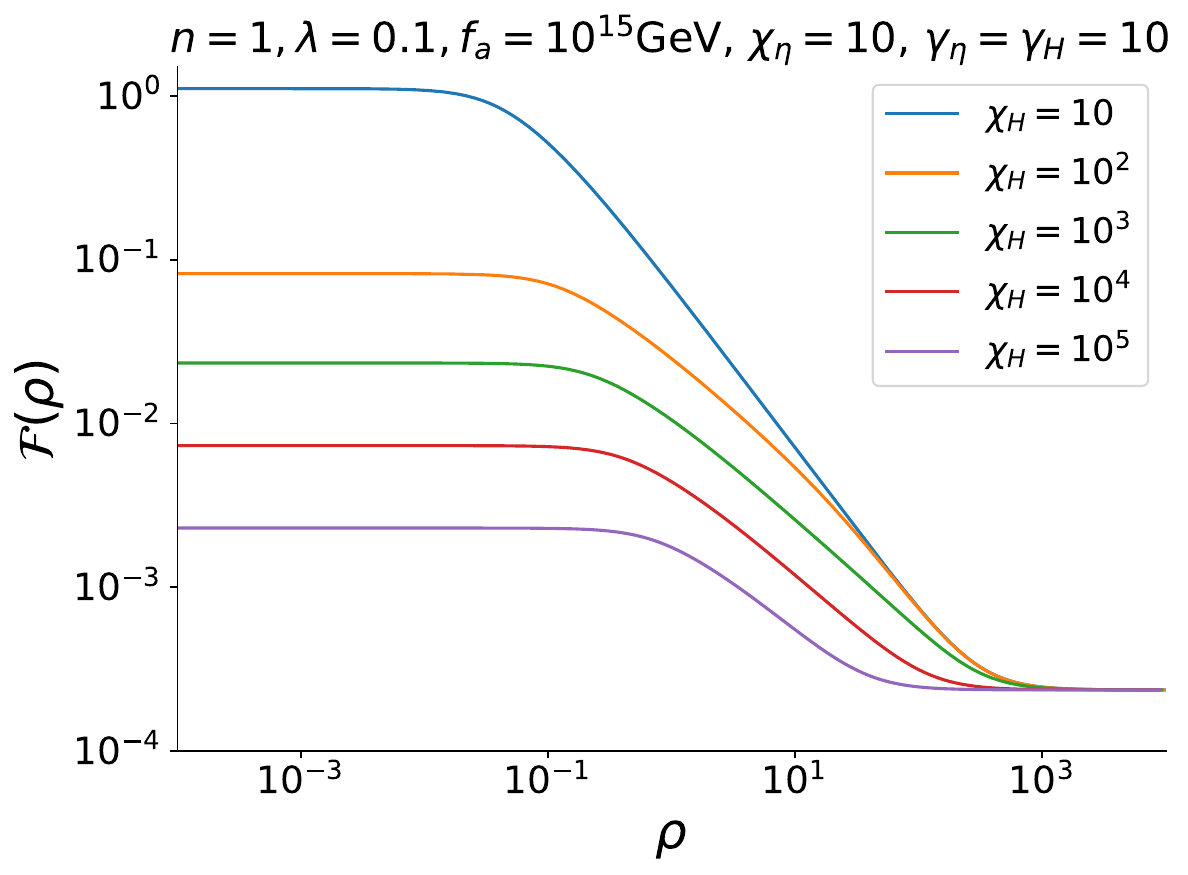}
  \hfill
  \includegraphics[clip, width=0.48\textwidth]{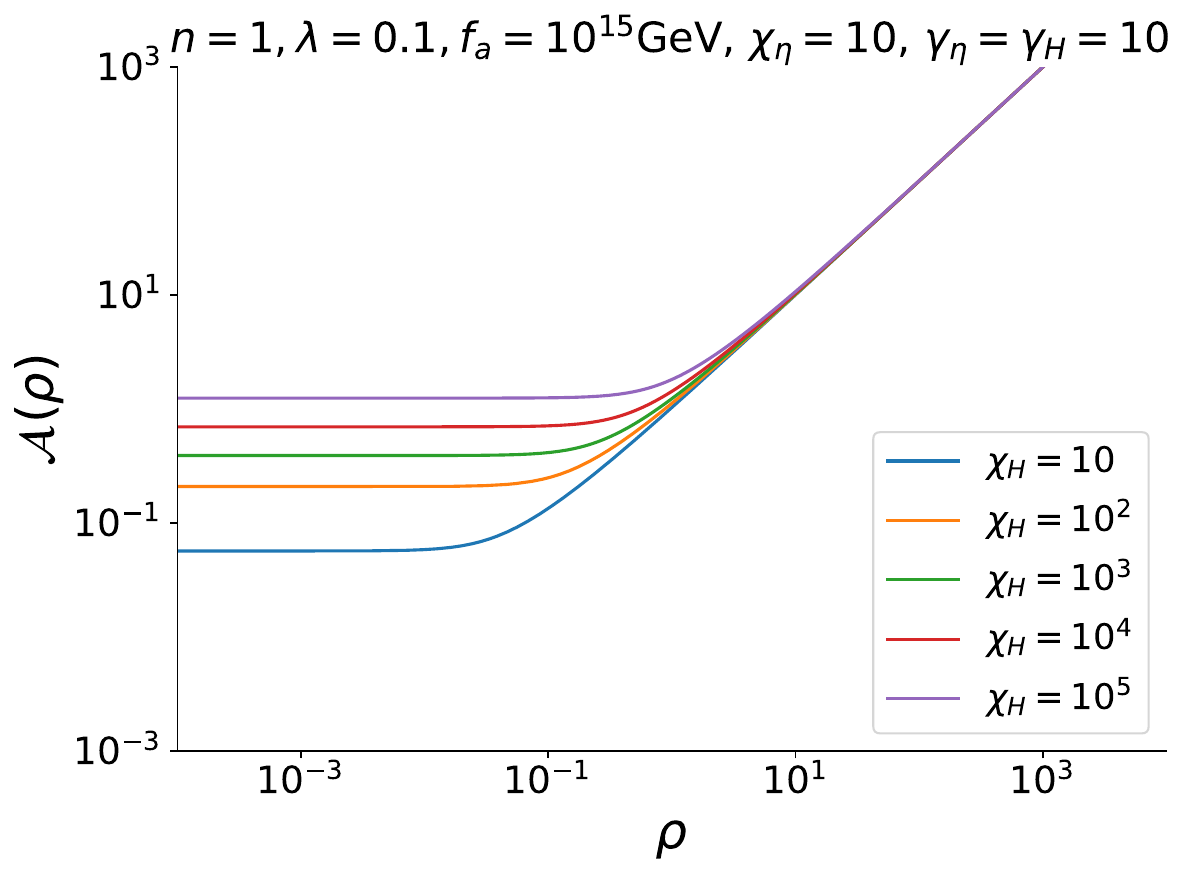}
  \caption{$\mathcal{F}(\rho)$ and $\mathcal{A}(\rho)$ for several values of $\chi_H = 10,~10^2,~10^3,~10^4$, and $10^5$ with $\chi_\eta =10$ and $\gamma_\eta = \gamma_H = 10$.}
  \label{fig:FA_Nieh_Holst}
\end{figure}

Fig.~\ref{fig:FA_Nieh_Holst} shows the axionic wormhole solutions of $\mathcal{F}(\rho)$ and $\mathcal{A}(\rho)$ for several values of $\chi_H$ with $\chi_\eta = 10$ and $\gamma_\eta = \gamma_H = 10$.
These results indicate that, as $\chi_H$ increases, the wormhole throat $\mathcal{A}(0)$ increases while the initial radial field value $\mathcal{F}(0)$ decreases.
Also, the profile for $\chi_\eta = \chi_H$, indicated by the solid blue curve, corresponds to the Palatini limit with $\xi=0$ because the factor $G$ becomes zero.

\begin{figure}[tbp]
  \centering
  \includegraphics[width=0.7\textwidth]{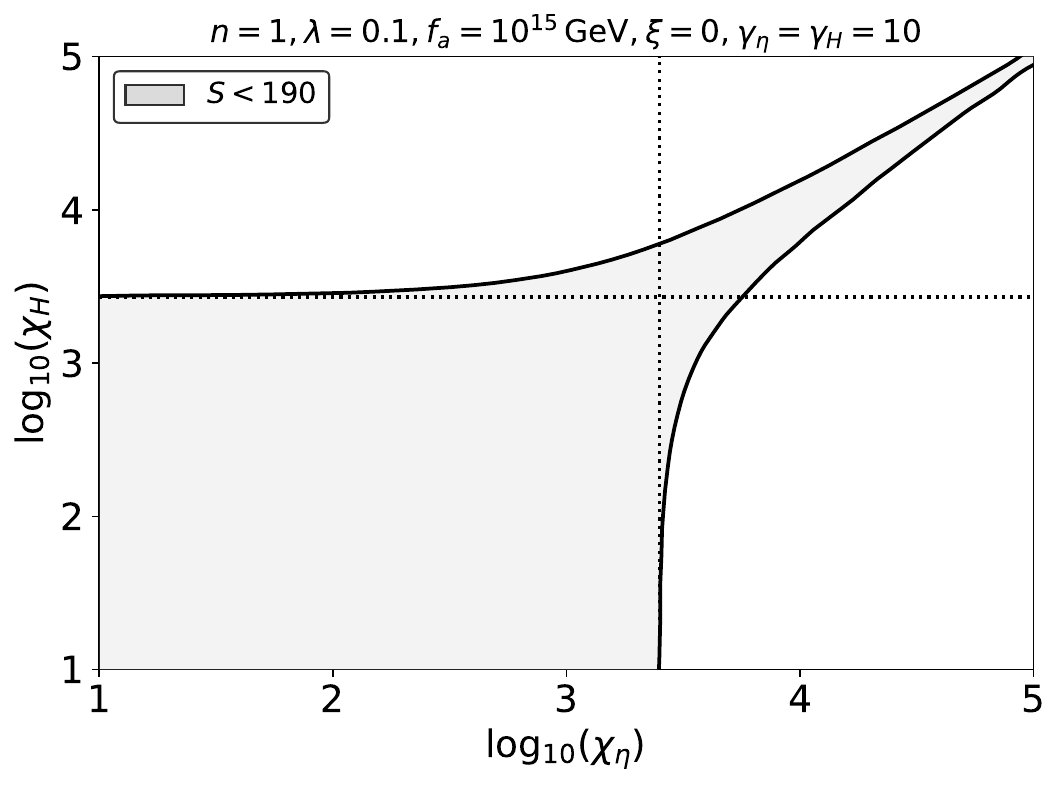}
\caption{
The region in the $(\chi_\eta,\chi_H)$ plane for $\xi = 0$ and $\gamma_\eta = \gamma_H = 10$. 
The gray shaded region corresponds to $S<190$, where the axion quality problem remains unsolved.
The black dotted lines indicate the thresholds required to solve the axion quality problem in each single-coupling limit.
Inflation cannot be successfully realized in any region.
}
\label{fig:WormInfNiehHolstXi0}
\end{figure}

In Fig.~\ref{fig:WormInfNiehHolstXi0}, we show the region in the $(\chi_\eta,\chi_H)$ plane where the axion quality problem remains unsolved, for fixed $\gamma_\eta = \gamma_H = 10$.
As mentioned above, the wormhole action depends on the two couplings mainly through the relative combination $(\chi_\eta - \chi_H)$.
We therefore find that the axion quality problem is resolved once $|\chi_\eta - \chi_H|$ becomes sufficiently large, and that the $S<190$ region forms a band around the cancellation line $\chi_\eta \simeq \chi_H$, where the Nieh--Yan and Holst contributions cancel each other.
The intersections of the $S<190$ boundary with the two axes are consistent with the corresponding single-coupling limits.
In this case, successful inflation cannot be realized in any region.

\subsection{Specific scenarios}
\label{sec:specific}

\begin{figure}[t]
  \centering
  \includegraphics[width=0.7\textwidth]{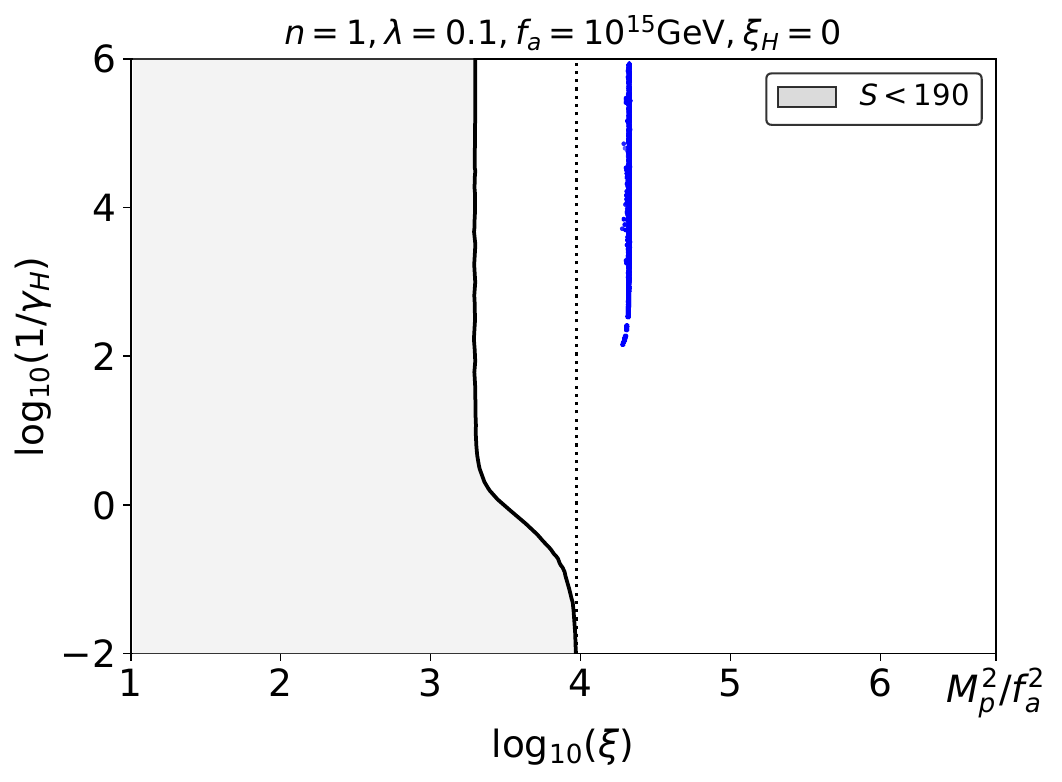}
\caption{
The parameter region in the $(\xi,1/\gamma_H)$ plane with $\xi_H=0$.
The gray shaded region corresponds to $S<190$, where the axion quality problem remains unsolved.
The black dotted line indicates the threshold $\xi^{\rm th}$.
The blue line denotes the inflation-compatible trajectory satisfying $N_e=80$, $0.97<n_s<0.98$, and $r<0.038$.
}
\label{fig:WormInfHolstGamma}
\end{figure}

In the previous subsections, we demonstrated that the axion quality problem can be resolved even in single-coupling cases, and explored broad parameter spaces involving combinations of non-minimal couplings.
However, we also found that the inflationary requirements severely restrict the regions where the axion quality problem is resolved. 
Indeed, both issues were simultaneously resolved only in the case of non-minimal couplings to the curvature scalar and the Nieh--Yan term.
In this subsection, we examine two additional parameter slices in order to clarify whether the viable parameter space for the axion quality problem can overlap with the regions compatible with inflation.

First, we consider the combination of the curvature coupling $\xi$ and the Barbero--Immirzi parameter $\gamma_H$.
This setup is directly motivated by the inflationary scenarios discussed in Refs.~\cite{Langvik:2020nrs, Shaposhnikov:2020gts}.
Fig.~\ref{fig:WormInfHolstGamma} shows the parameter region in the $(\xi,1/\gamma_H)$ plane with $\xi_H=0$.
We find that the axion quality problem is resolved for $\xi \gtrsim 2.0 \times 10^3$ in the small $\gamma_H$ regime.
Since the dynamics of the axionic wormhole can be approximately described by non-minimal coupling to the curvature scalar in the large $\gamma_H$ region, the required value of $\xi$ increases toward $\xi^{\rm th}$ as $\gamma_H$ becomes large, as anticipated in Sec.~\ref{sec:curvatureHolst}.
Furthermore, this setup simultaneously accommodates successful inflation only along the blue line $\xi \sim 2.0 \times 10^4$ within the range $1/\gamma_H \gtrsim 1.4 \times 10^2$.
Note that inflation can be successfully realized up to $1 / \gamma_H \simeq 10^6$, as discussed in Appendix~\ref{sec:inflation}.

\begin{figure}[t]
  \centering
  \includegraphics[width=0.7\textwidth]{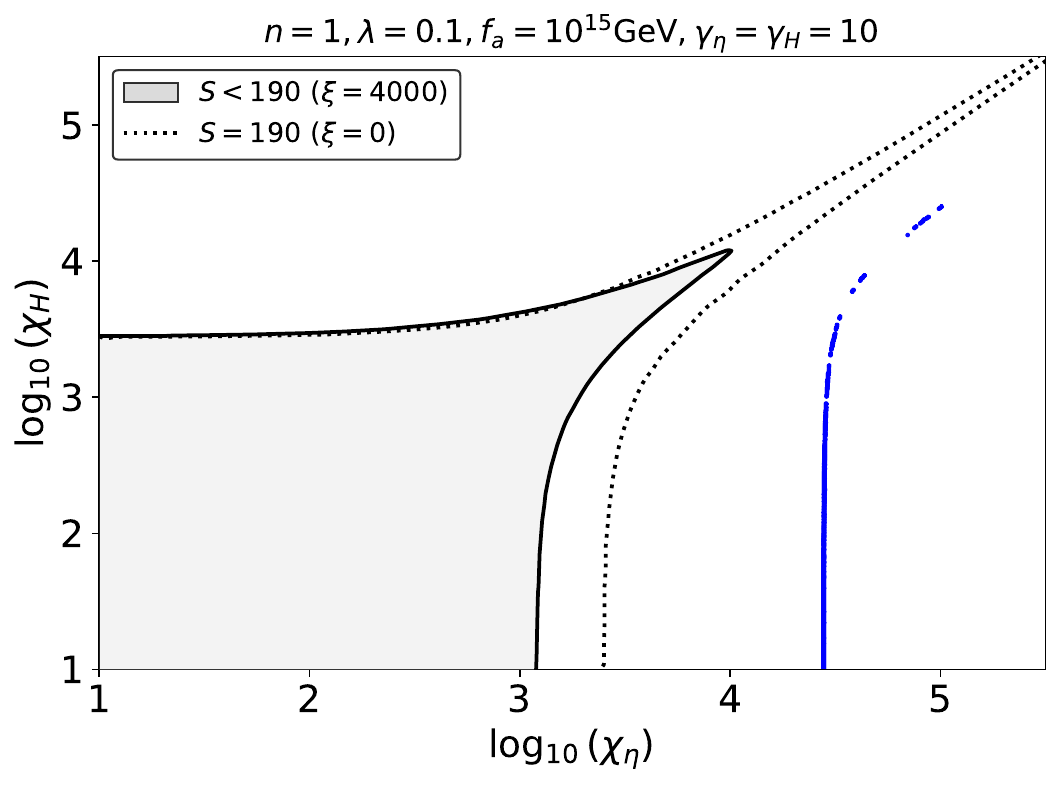}
\caption{
The parameter region in the $(\chi_\eta,\chi_H)$ plane for $\xi=0$ and $4000$ with $\gamma_\eta=\gamma_H=10$.
The gray shaded region corresponds to $S<190$ for $\xi=4000$, where the axion quality problem remains unsolved, while the black dotted line shows the boundary $S = 190$ for $\xi = 0$.
The blue curve denotes the inflation-compatible trajectory satisfying $N_e=80$, $0.97<n_s<0.98$, and $r<0.038$ in the case of $\xi = 4000$.
}
\label{fig:WormInfNiehHolstXi4000}
\end{figure}

Second, we revisit the case of non-minimal couplings to the Nieh--Yan and Holst terms by introducing a finite curvature coupling $\xi$, which can change the inflationary predictions.
Fig.~\ref{fig:WormInfNiehHolstXi4000} shows the parameter region in the $(\chi_\eta,\chi_H)$ plane for $\xi=0$ and $4000$ with $\gamma_\eta=\gamma_H=10$.
The choice of $\xi=4000$ is motivated by the inflationary analysis summarized in Appendix~\ref{sec:inflation}.
We find that the axion quality problem is resolved in a broad region, except near $\chi_\eta \simeq \chi_H$, where the contributions of the Nieh--Yan and Holst terms nearly cancel each other.
Compared with the $\xi=0$ case, this cancellation region along $\chi_\eta \simeq \chi_H$ no longer extends indefinitely; it closes near $\chi_\eta \simeq 1.1 \times 10^4$ and $\chi_H \simeq 1.3 \times 10^4$.
This is because, in the large $\chi_H$ region, the dynamics of the axionic wormhole can be approximately described by non-minimal couplings to the curvature scalar and the Holst term.
The intersection of the $S<190$ boundary with the $\chi_H$ axis is close to the threshold $\chi_H^{\rm th}$, while the intersection with the $\chi_\eta$ axis lies below the threshold $\chi_\eta^{\rm th}$.
This behavior is consistent with the single-coupling analyses at $\xi = 4000$: in the $(\xi,\chi_H)$ plane of Fig.~\ref{fig:WormInfHolst}, the $S=190$ boundary at $\xi=4000$ nearly coincides with $\chi_H^{\rm th}$, whereas in the $(\xi,\chi_\eta)$ plane of Fig.~\ref{fig:WormInfNieh}, the $S=190$ boundary at $\xi=4000$ lies below $\chi_\eta^{\rm th}$.
The region compatible with inflation overlaps with the viable parameter space for the axion quality problem only along the blue curve, as shown in Fig.~\ref{fig:WormInfNiehHolstXi4000}.

\section{Conclusions and Discussion}
\label{sec:Conclusion}

In this paper, we have studied axionic wormholes in Metric-Affine Gravity (MAG).
MAG allows for spacetime torsion and non-metricity, and therefore admits additional curvature-like and topological terms such as the Holst and Nieh--Yan terms.
After solving for the independent affine connection, these terms affect the wormhole equations through the factor $G$.
This factor modifies the effective radial dynamics of the PQ field and provides a mechanism for enhancing the Euclidean wormhole action.
As a result, the wormhole-induced PQ violation can be sufficiently suppressed in parameter regions that have no counterpart in the metric and Palatini formalisms.

We first analyzed single-coupling cases.
The case where only the curvature scalar is non-minimally coupled reproduces the result in the Palatini formalism, where the axion quality problem can be alleviated for $\xi \gtrsim 9.4 \times 10^{3}$ \cite{Cheong:2022ikv}.
In the case where only the Nieh--Yan term is non-minimally coupled, the action and the wormhole dynamics are controlled by $\chi_\eta = \xi_\eta/\gamma_\eta$, and the axion quality problem can be alleviated for $\chi_\eta \gtrsim 2.5 \times 10^{3}$.
In the case where only the Holst term is non-minimally coupled, the character of the wormhole dynamics depends on the size of the Barbero--Immirzi parameter $\gamma_{H}$.
The dynamics is controlled by $\chi_H = \xi_H/\gamma_H$ in the large $\gamma_H$ regime.
The axion quality problem can be alleviated for $\chi_H \gtrsim 2.7 \times 10^3$ in this regime.
In the small $\gamma_H$ regime, the dynamics is instead governed by $\xi_H$, so that $\chi_H$ must be larger.

Next, we analyzed two-coupling cases: the curvature scalar and the Nieh--Yan term, the curvature scalar and the Holst term, and the Nieh--Yan and Holst terms.
We found that, in all cases, the viable parameter space is enlarged relative to that in the corresponding single-coupling cases.
A characteristic cancellation structure appears in the latter two combinations.
For the curvature scalar and the Holst term, the factor $G$ is proportional to $(\xi-\xi_H)^2$, and the theory approaches the Palatini limit near the cancellation line $\xi\simeq\xi_H$.
For the Nieh--Yan and Holst terms, the factor $G$ is proportional to $(\chi_\eta-\chi_H)^2$, and, near the cancellation line $\chi_\eta\simeq\chi_H$, the theory approaches the Palatini limit with the minimal coupling to the curvature scalar.
Furthermore, we explored regions that can simultaneously accommodate successful inflation using the benchmark conditions $N_e=80$,~$0.97<n_s<0.98$, and $r<0.038$.
Since the inflationary requirements are satisfied only by the non-minimal couplings to the curvature scalar and the Nieh--Yan term,
we further investigated viable overlapping regions in two representative cases: the curvature scalar coupling $\xi$ combined with the Barbero--Immirzi parameter $\gamma_H$, and the Nieh--Yan and Holst terms with a finite curvature scalar coupling $\xi$.

A complete analysis of perturbative unitarity is beyond the scope of the present work.
Nevertheless, it is an important consistency condition because the relevant parameter regions involve large non-minimal couplings.
As is known from Higgs inflation, the perturbative cutoff can depend on the gravitational formulation, the background field configuration, and even the scalar self-coupling through higher-point scattering amplitudes~\cite{Barbon:2009ya,Bauer:2010jg,Ito:2021ssc,Steingasser:2025txd,He:2026fzs}.
In MAG, the factor $G$ modifies the scalar kinetic structure, so the cutoff may also depend on the non-minimal couplings and on the background field configuration~\cite{Shaposhnikov:2020gts}.
We leave a dedicated analysis of this issue for future work.

Our results also suggest further cosmological applications.
Axionic wormholes induce PQ-breaking operators whose coefficients are exponentially suppressed by $e^{-S}$, and thus the MAG mechanism found here may also be relevant for wormhole-induced axion-like particle~(ALP) dark matter scenarios~\cite{Cheong:2024kum}.
We note that, in post-inflationary scenarios, even explicit PQ breaking small enough to satisfy the strong CP constraint can affect the axion abundance through its influence on the early dynamics of the axion field~\cite{Zantedeschi:2026iql}.
It would be interesting to extend this analysis to a more complete cosmological setting including inflation, reheating, and ALP dark matter phenomenology, together with the consistency conditions discussed above.

\section*{Acknowledgements}
ST thanks Ryosuke Sato and Wen Yin for useful discussions.
The work of ST is supported by JST SPRING, Grant Number JPMJSP2132.
NY thanks Tomohiro Inagaki for useful discussions.

We acknowledge the use of OpenAI's ChatGPT and Anthropic's Claude Code for language editing of the manuscript and to assist in the development of numerical code used in the calculations. All AI-assisted outputs were independently checked and revised by the authors, who take full responsibility for the scientific content and results presented in this paper.

\appendix

\section{The behaviors of ${\mathcal F}(0)$ and ${\mathcal A}(0)$}
\label{sec:F0A0}

\begin{figure}[!htbp]
  \centering
  \includegraphics[width=0.58\textwidth]{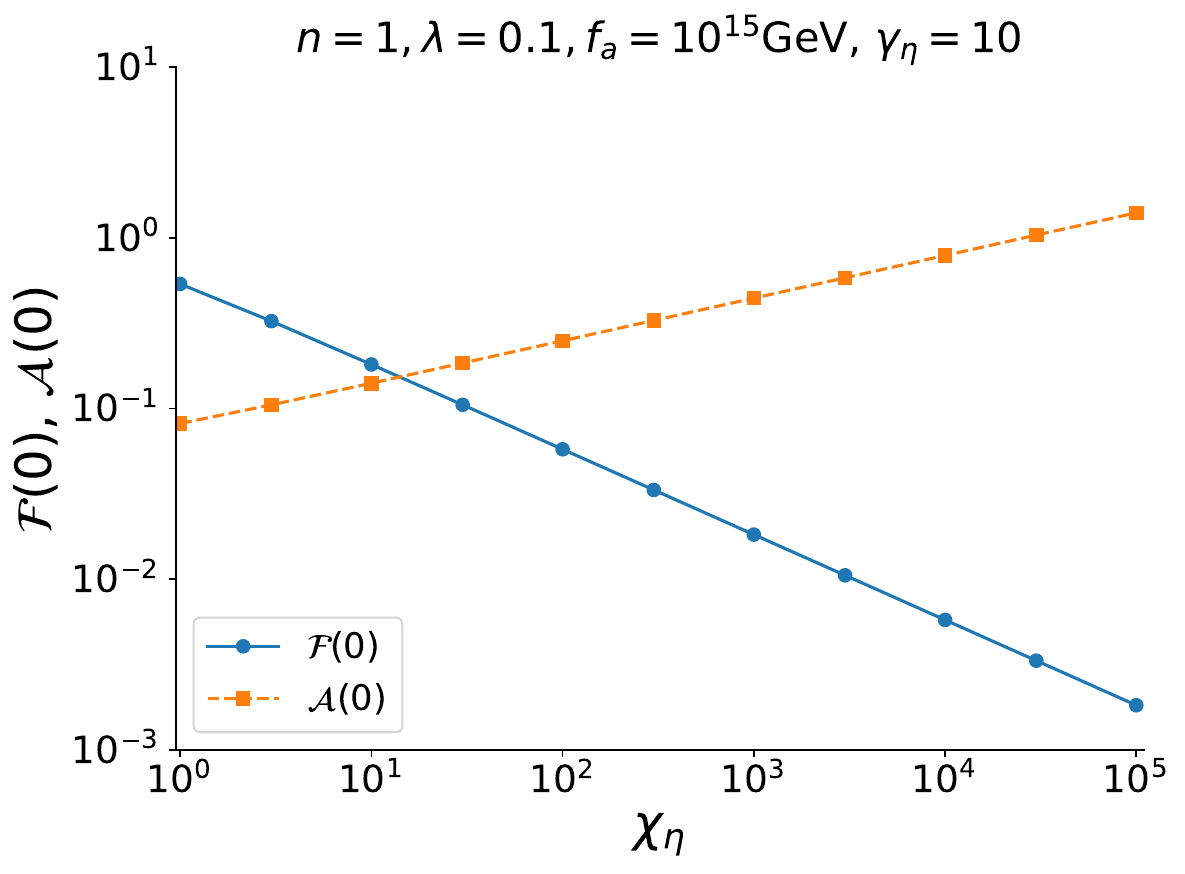}
\caption{
The behaviors of ${\mathcal F}(0)$ and ${\mathcal A}(0)$ in the case with only the Nieh--Yan term.
}
\label{fig:F0A0_NH}
\end{figure}

\begin{figure}[t]
  \centering
  \includegraphics[clip, width=0.48\textwidth]{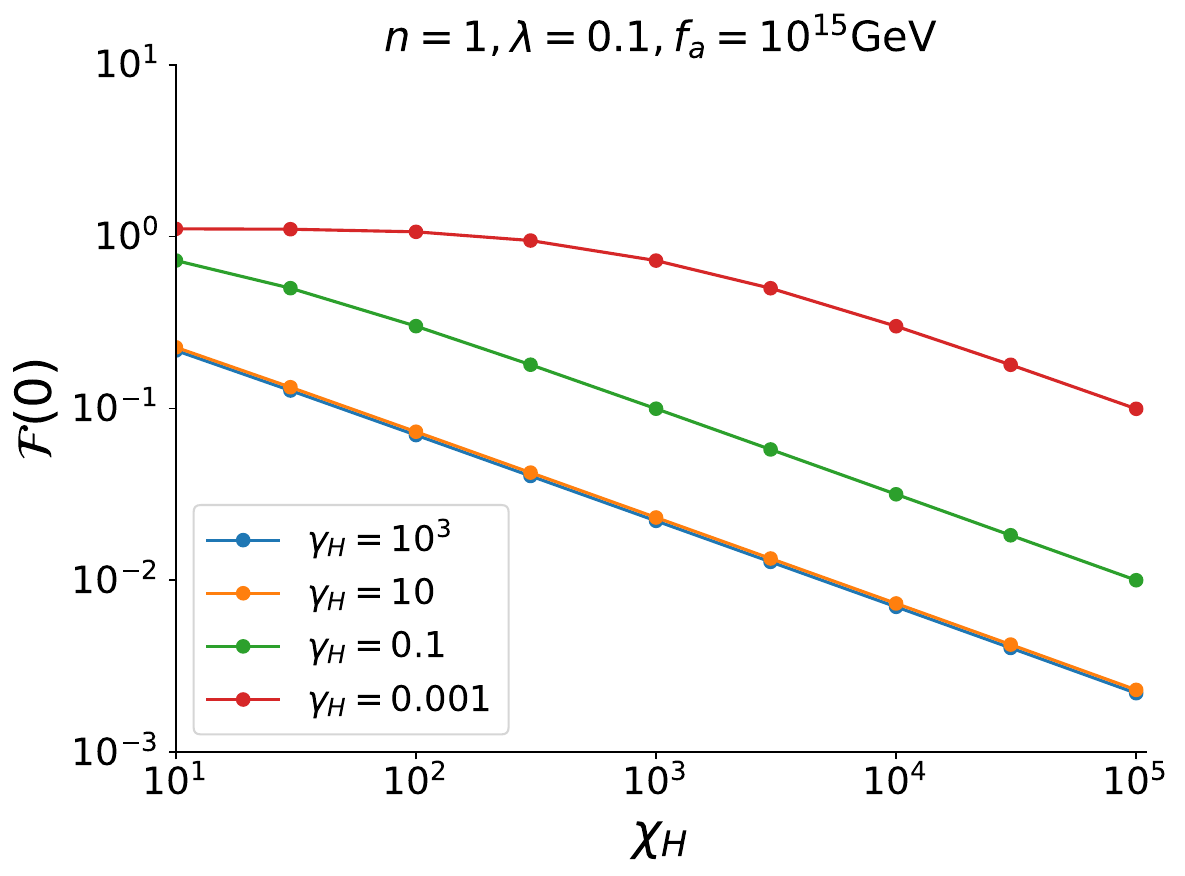}
  \hfill
  \includegraphics[clip, width=0.48\textwidth]{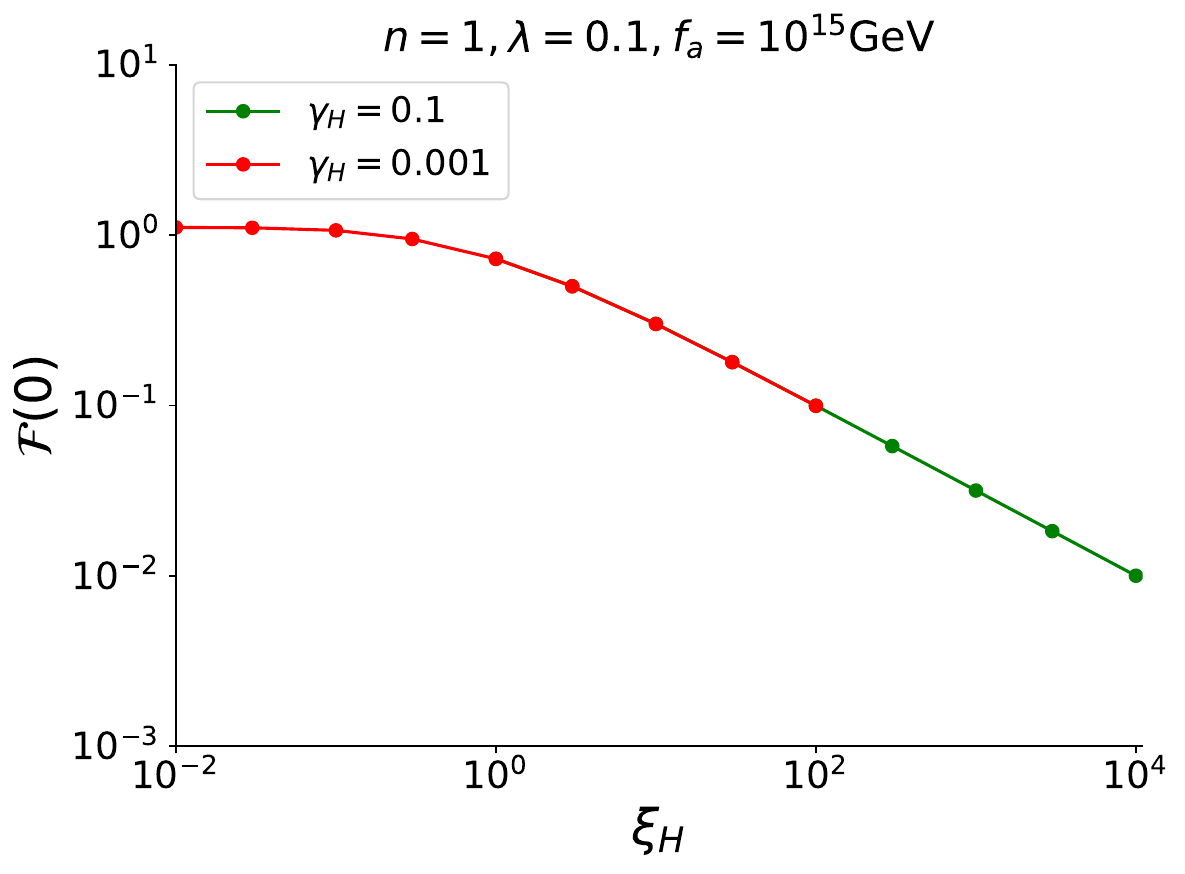}
  \includegraphics[clip, width=0.48\textwidth]{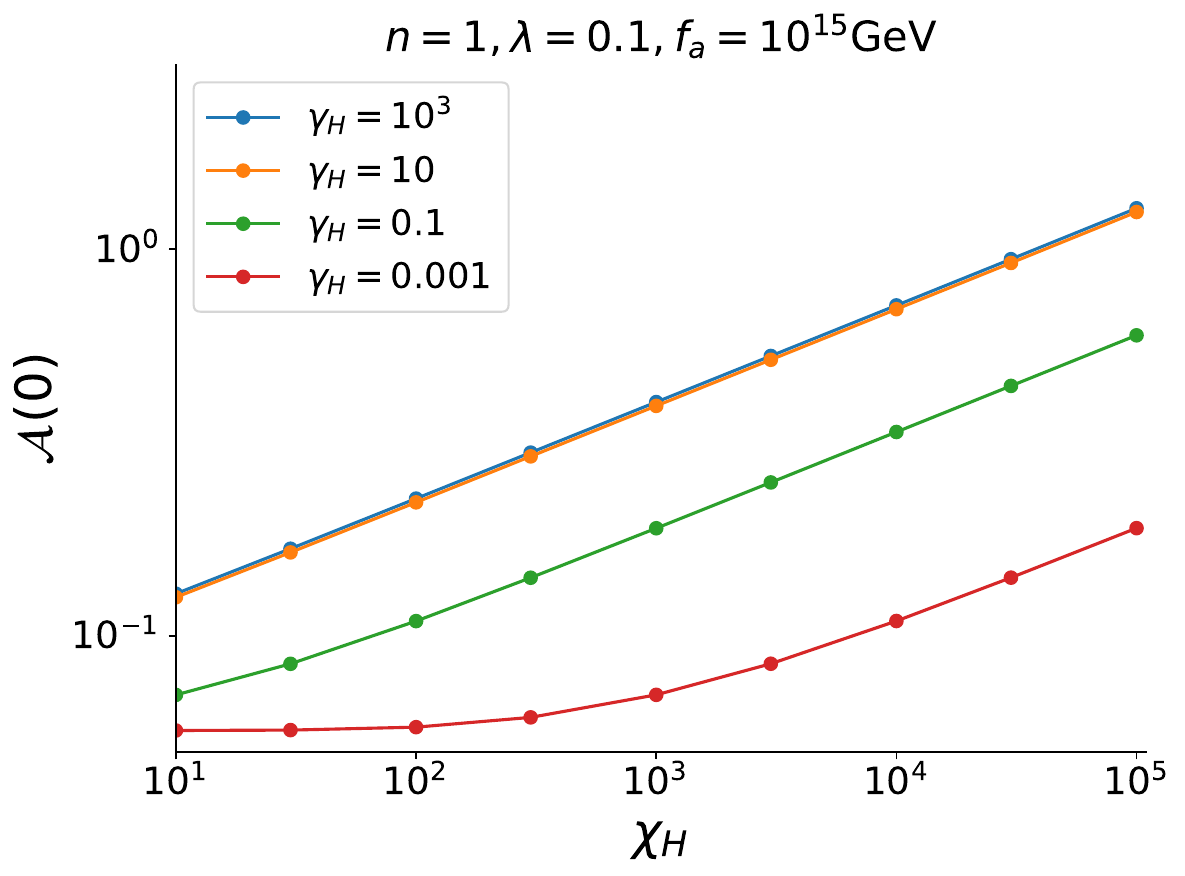}
  \hfill
  \includegraphics[clip, width=0.48\textwidth]{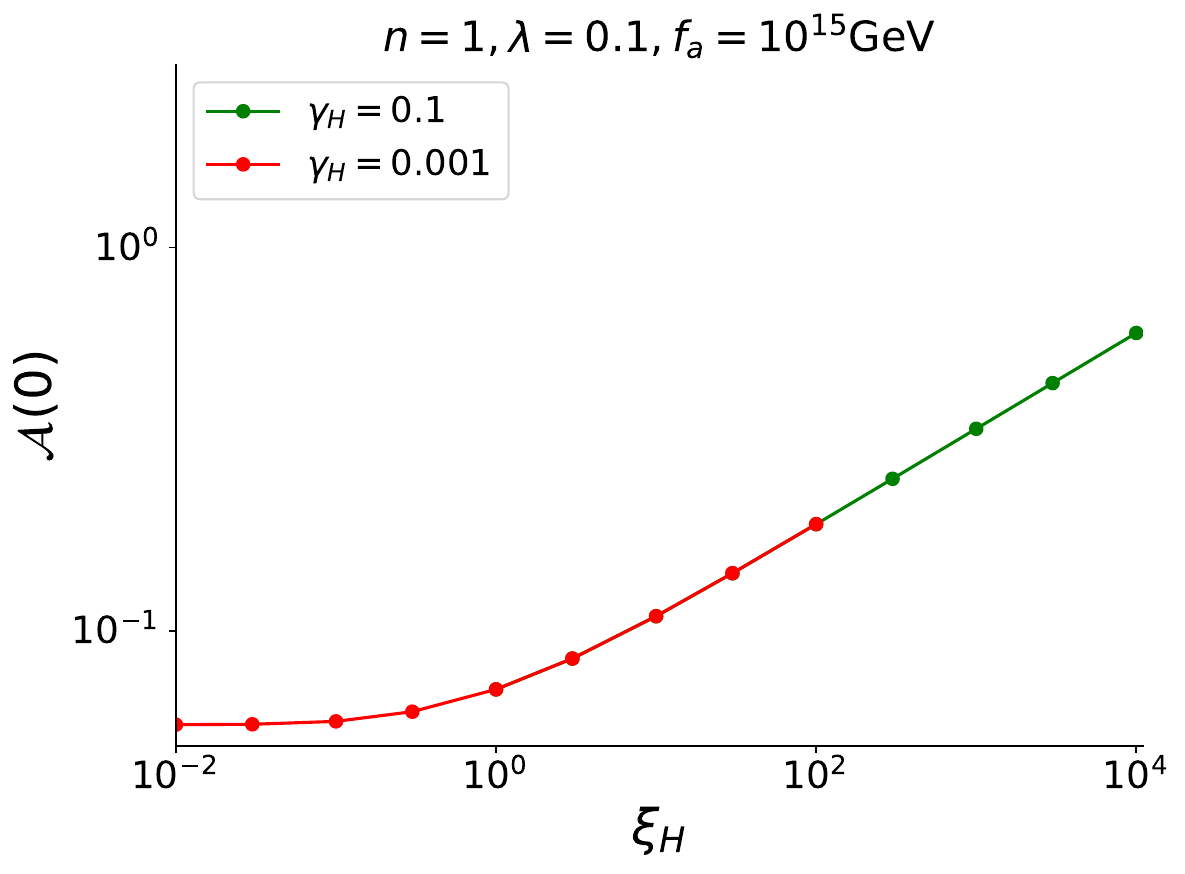}
  \caption{The behaviors of ${\mathcal F}(0)$ and ${\mathcal A}(0)$ in the case with only the Holst term.}
  \label{fig:F0A0_Holst}
\end{figure}

In this appendix, we examine the profiles of wormhole solutions discussed in Sec.~\ref{sec:NA} by showing the behaviors of the initial radial field value ${\mathcal F}(0)$ and the wormhole throat ${\mathcal A}(0)$.

First, we discuss single-coupling cases.
In the case with only the curvature scalar, which corresponds to the Palatini formalism, the axionic wormhole solutions were investigated in Ref.~\cite{Cheong:2022ikv}.
As $\xi$ increases, ${\mathcal F}(0)$ takes a value of $\mathcal{O}(1)$ except for values extremely close to $\xi = M_p^2 / f_a^2$.
On the other hand, ${\mathcal A}(0)$ initially decreases and then increases drastically near $\xi = M_p^2 / f_a^2$.
For $\xi = M_p^2 / f_a^2$, this solution is identical to the Giddings--Strominger wormhole~\cite{Giddings:1987cg}.

Fig.~\ref{fig:F0A0_NH} shows the behaviors of ${\mathcal F}(0)$ and ${\mathcal A}(0)$ for the case where the Nieh--Yan term is non-minimally coupled with $\gamma_\eta$ fixed to $10$.
In this case, as can be seen from Eq.~\eqref{eq:G_Nieh}, the dynamics of the axionic wormhole depends only on $\chi_\eta$.
We find that $\mathcal{F}(0)$ decreases and $\mathcal{A}(0)$ increases as $\chi_\eta$ increases.
Fig.~\ref{fig:F0A0_Holst} shows the behaviors of ${\mathcal F}(0)$ and ${\mathcal A}(0)$ for the case where the Holst term is non-minimally coupled.
The blue, orange, green, and red lines indicate the results for $\gamma_H = 10^3,~10,~0.1$, and $10^{-3}$, respectively.
As seen from Eqs.~\eqref{eq:G_Holst_large} and \eqref{eq:G_Holst_small}, the dynamics of the axionic wormhole depends on $\chi_H$ for large $\gamma_H$, and on $\xi_H$ for small $\gamma_H$.
We find that the behavior of the axionic wormhole is similar to that in the case with only the Nieh--Yan term.
Indeed, ${\mathcal F}(0)$ decreases and ${\mathcal A}(0)$ increases as $\chi_H~(\xi_H)$ increases.

\begin{figure}[h]
  \centering
  \includegraphics[clip, width=0.48\textwidth]{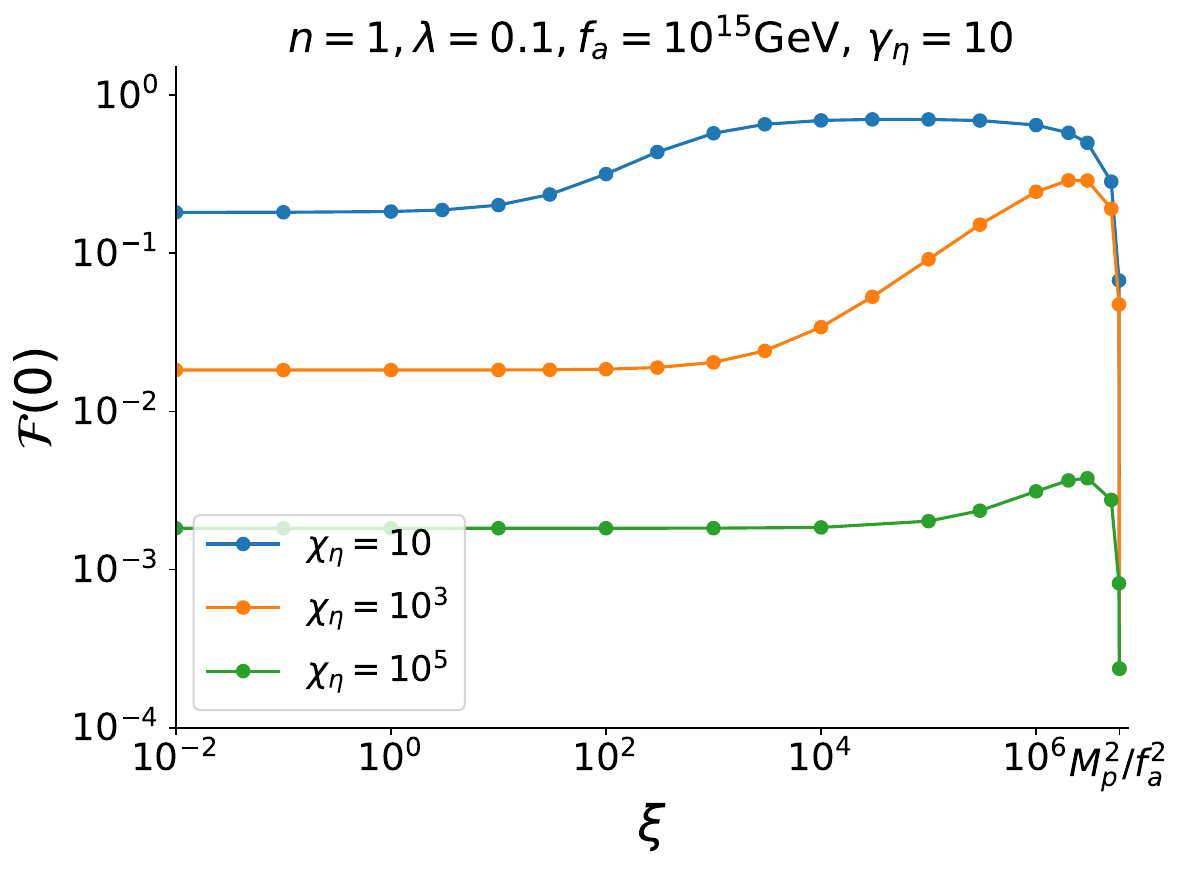}
  \hfill
  \includegraphics[clip, width=0.48\textwidth]{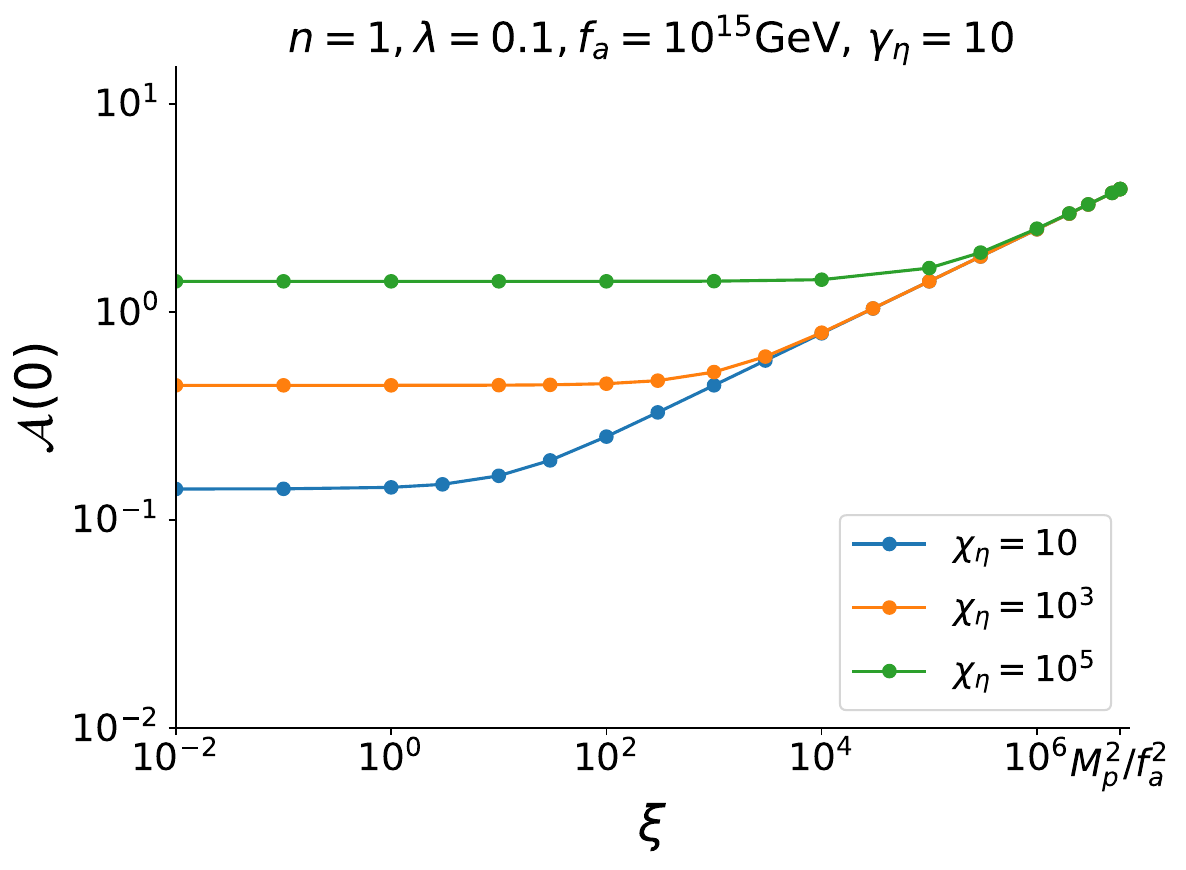}
  \caption{
  The behaviors of ${\mathcal F}(0)$ and ${\mathcal A}(0)$ in the case of the curvature scalar and the Nieh--Yan term.
  }
  \label{fig:F0A0_nonminimal_Nieh}
\end{figure}

Next, we turn to two-coupling cases.
First, we consider the combination of the curvature scalar and the Nieh--Yan term.
Fig.~\ref{fig:F0A0_nonminimal_Nieh} shows the behavior of ${\mathcal F}(0)$ and ${\mathcal A}(0)$ as functions of $\xi$ for various values of $\chi_\eta$ at fixed $\gamma_\eta = 10$.
From Eq.~\eqref{eq:G_nonminimal_Nieh}, for large $\xi$, the factor $G$ mostly scales as $\chi_\eta^2 / \xi$, while for small $\xi$, it mostly depends on $\chi_\eta^2$.
We find that in the small $\xi$ region, the behavior of the axionic wormhole is similar to that of the case with only the Nieh--Yan term, in which $\mathcal{F}(0)$ decreases and $\mathcal{A}(0)$ increases as $\chi_\eta$ increases. 
On the other hand, in the large $\xi$ region, the factor $G$ is strongly suppressed by $\xi$, and the solutions approach those of the Palatini formalism.
For $\xi = M_p^2 / f_a^2$, this solution is identical to the Giddings--Strominger wormhole~\cite{Giddings:1987cg}.

\begin{figure}[t]
  \centering
  \includegraphics[clip, width=0.48\textwidth]{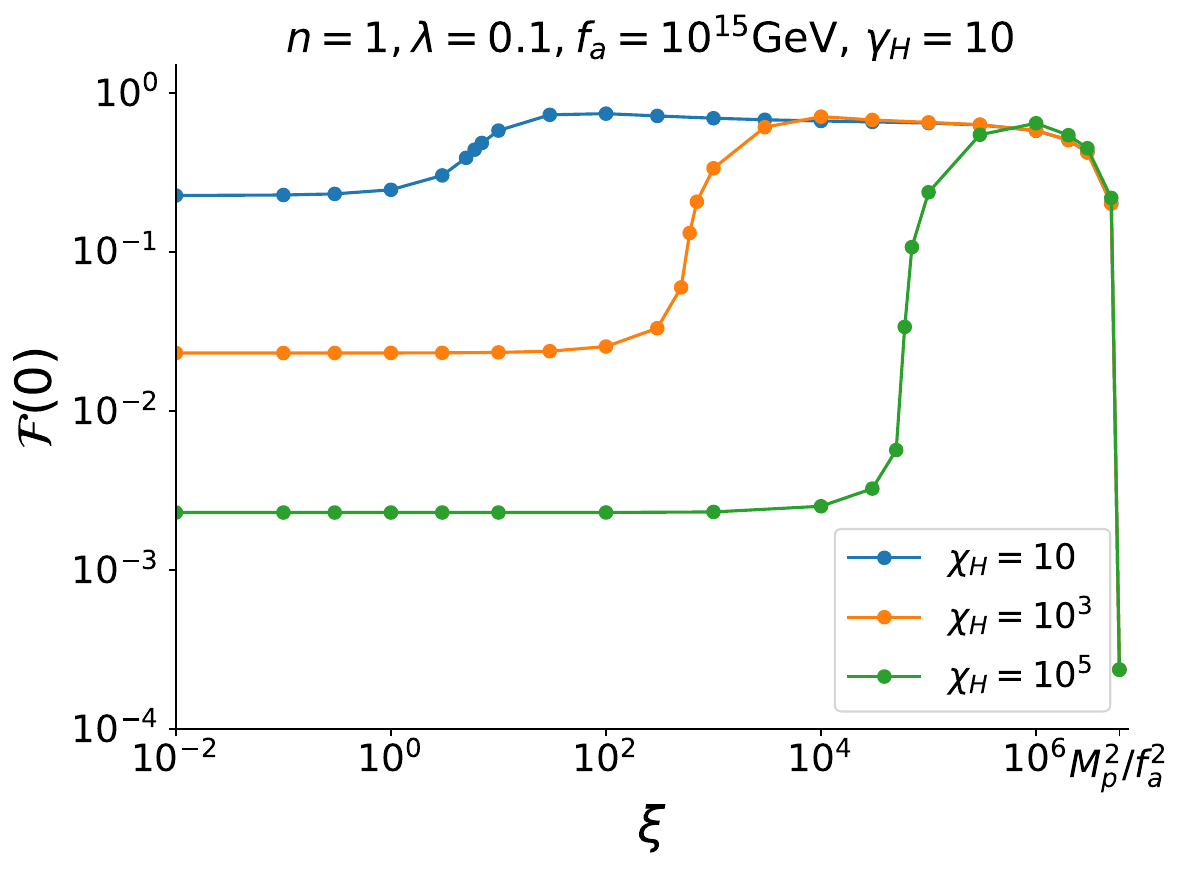}
  \hfill
  \includegraphics[clip, width=0.48\textwidth]{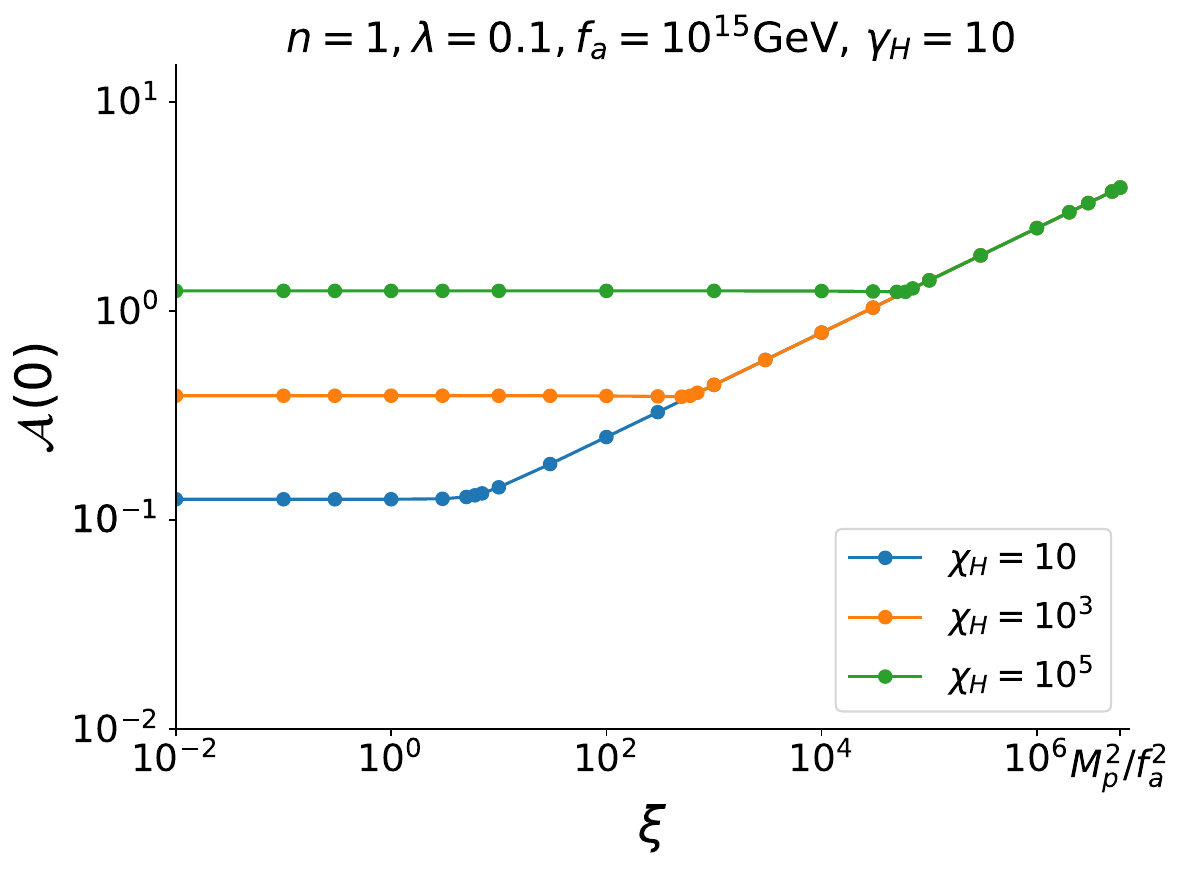}
  \caption{The behaviors of ${\mathcal F}(0)$ and ${\mathcal A}(0)$ in the case of the curvature scalar and the Holst term with $\gamma_H = 10$.
  }
  \label{fig:F0A0_nonminimal_Holst_10}
\end{figure}

\begin{figure}[t]
  \centering
  \includegraphics[clip, width=0.48\textwidth]{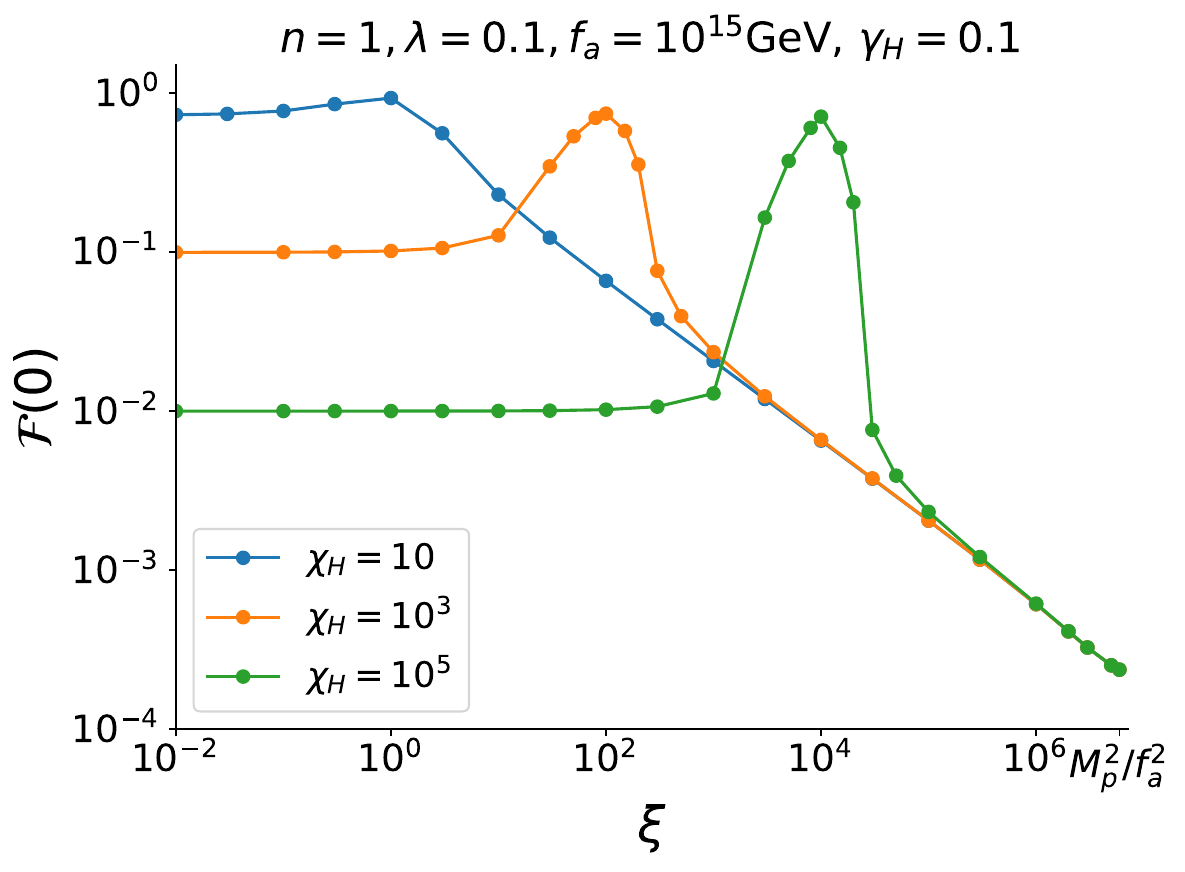}
  \hfill
  \includegraphics[clip, width=0.48\textwidth]{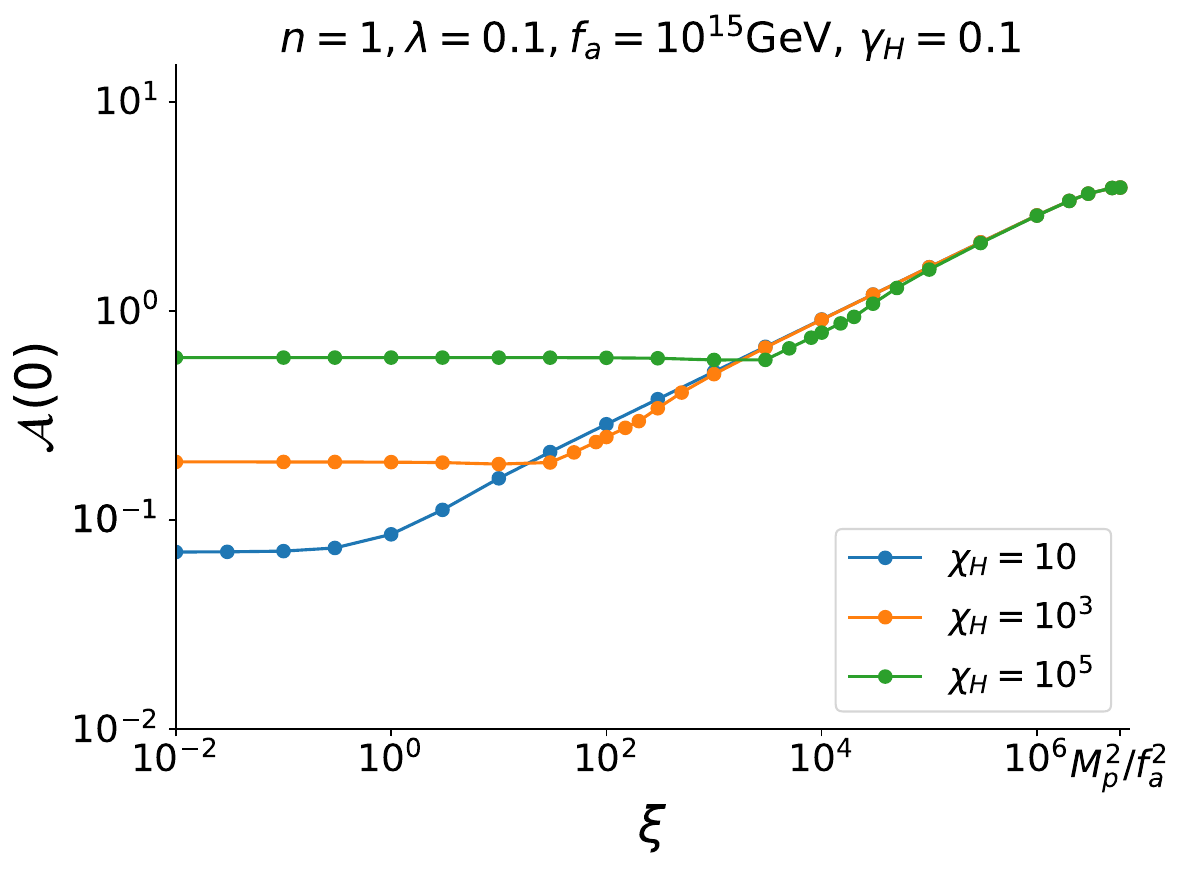}
  \caption{The behaviors of ${\mathcal F}(0)$ and ${\mathcal A}(0)$ in the case of the curvature scalar and the Holst term with $\gamma_H = 0.1$.
  }
  \label{fig:F0A0_nonminimal_Holst_01}
\end{figure}

Next, we consider the combination of the curvature scalar and the Holst term.
In this case, the wormhole dynamics is characterized by $\gamma_H$ and $\xi$, as shown in Eq.~\eqref{eq:factorGXiHolst}.
For large $\gamma_H$, the $\gamma_H^2(1 + 3\xi(\mathcal{F}^2 - \mathcal{F}_a^2))^2$ term dominates the bracket in the denominator.
In the large $\xi$ region, the factors of $(1 + 3\xi(\mathcal{F}^2 - \mathcal{F}_a^2))$ in the numerator and denominator largely cancel, so that $G$ is suppressed as $G \sim 1/ (\xi \gamma_H^2 \mathcal{F}^4)$ and approaches the Palatini limit. 
In the small $\xi$ region, $G$ mostly depends on $\chi_H^2$ as $G \sim 18\chi_H^2\mathcal{F}^2$.
For small $\gamma_H$, the same bracket is instead dominated by the $(1 + 3\xi_H(\mathcal{F}^2 - \mathcal{F}_a^2))^2$ term.
Then, $G$ depends on $\xi$ for large $\xi$ and on $\xi_H$ for small $\xi$, as in the small $\gamma_H$ regime of the case with only the Holst term.
Figs.~\ref{fig:F0A0_nonminimal_Holst_10} and \ref{fig:F0A0_nonminimal_Holst_01} show the behavior of ${\mathcal F}(0)$ and ${\mathcal A}(0)$ as functions of $\xi$ for various values of $\chi_H$, with $\gamma_H$ fixed to $10$ and $0.1$, respectively.
We find that in the small $\xi$ region, ${\mathcal F} (0)$ and ${\mathcal A} (0)$ depend only on $\chi_H~(\xi_H)$.
In the large $\xi$ region, the factor $G$ is strongly suppressed by $\gamma_H$ for large $\gamma_H$, and the solutions approach those of the Palatini formalism.
On the other hand, for small $\gamma_H$ and large $\xi$, the factor $G$ primarily depends on $\xi$, and both $\mathcal{F}(0)$ and $\mathcal{A}(0)$ asymptotically converge to their respective universal values at large $\xi$ regardless of the initial difference in $\chi_H$.
For $\xi = M_p^2 / f_a^2$, this solution is identical to the Giddings--Strominger wormhole~\cite{Giddings:1987cg}.

\begin{figure}[t]
  \centering
  \includegraphics[clip, width=0.48\textwidth]{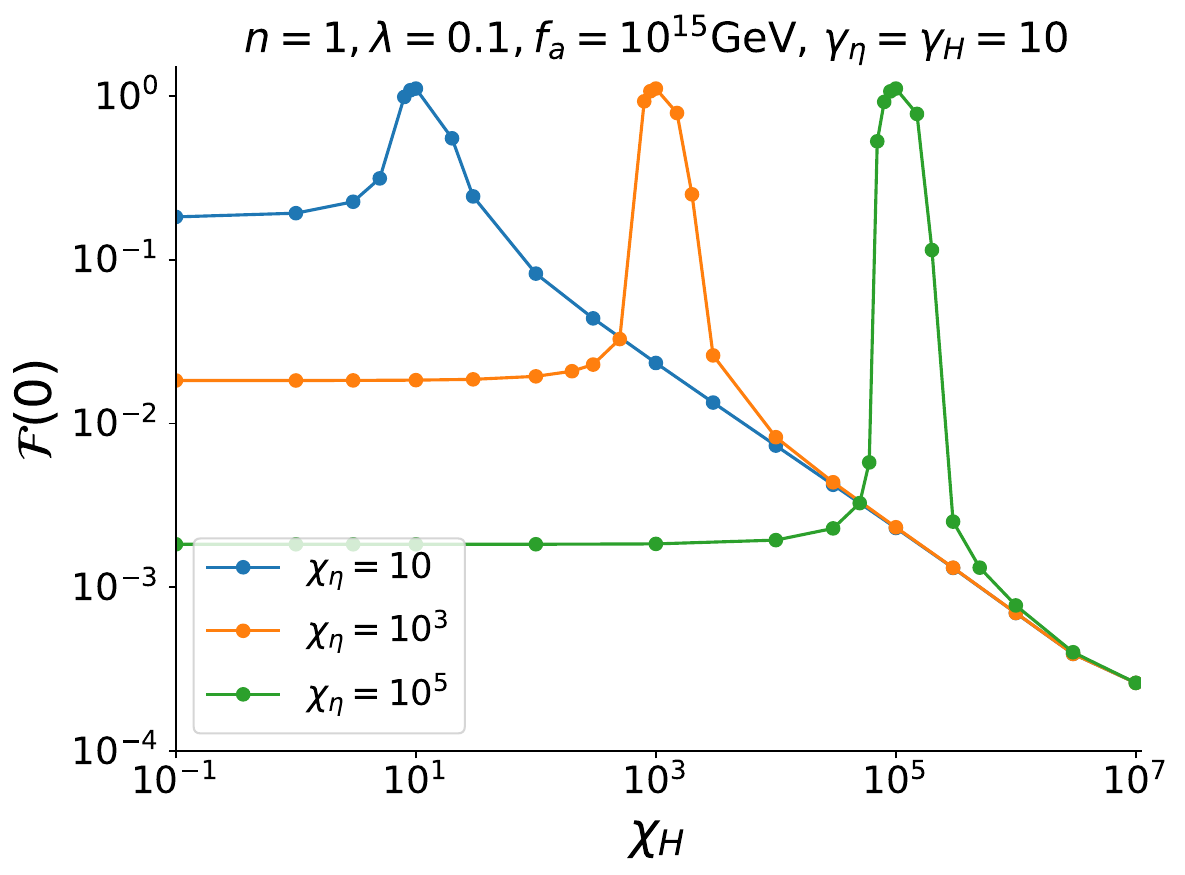}
  \hfill
  \includegraphics[clip, width=0.48\textwidth]{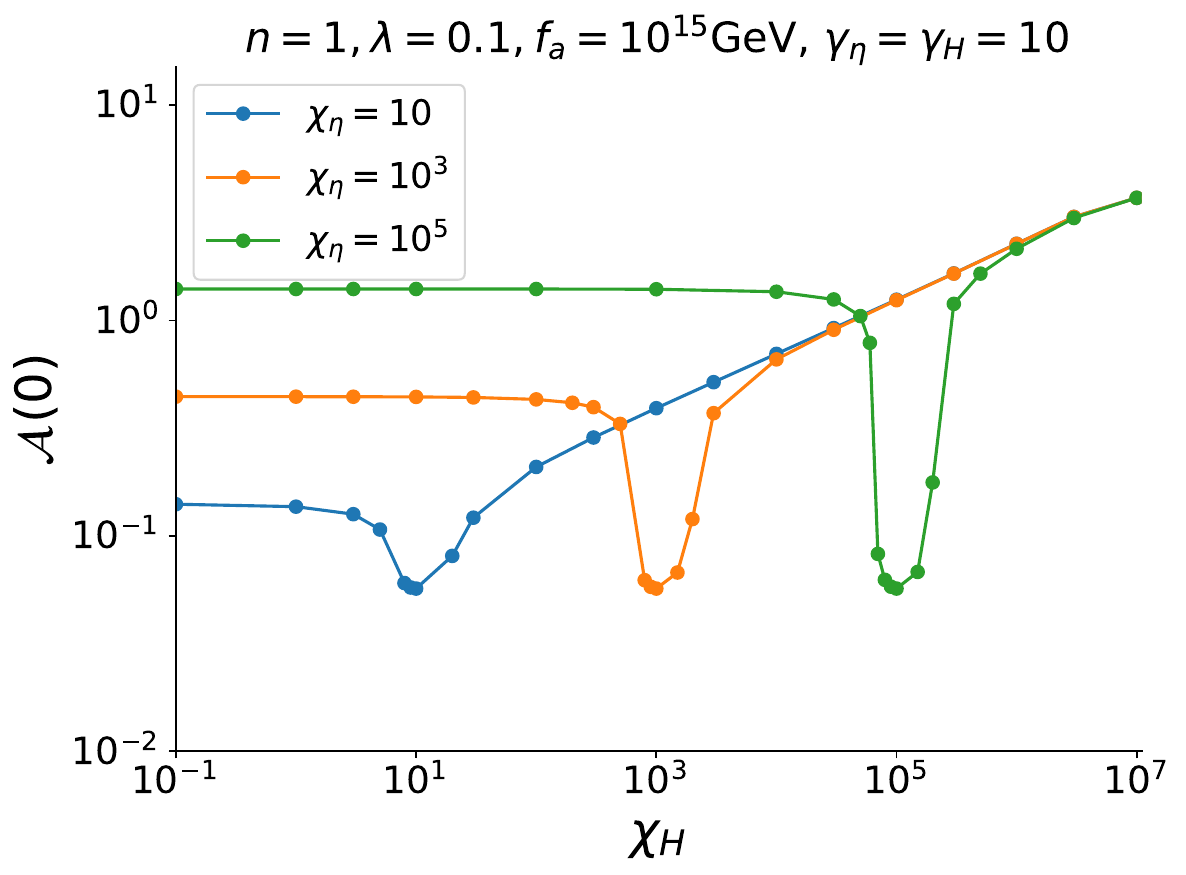}
  \caption{The behaviors of ${\mathcal F}(0)$ and ${\mathcal A}(0)$ in the case of the Nieh--Yan and Holst terms with $\gamma_\eta = \gamma_H = 10$.}
  \label{fig:F0A0_Nieh_Holst}
\end{figure}

\begin{figure}[t]
  \centering
  \includegraphics[clip, width=0.48\textwidth]{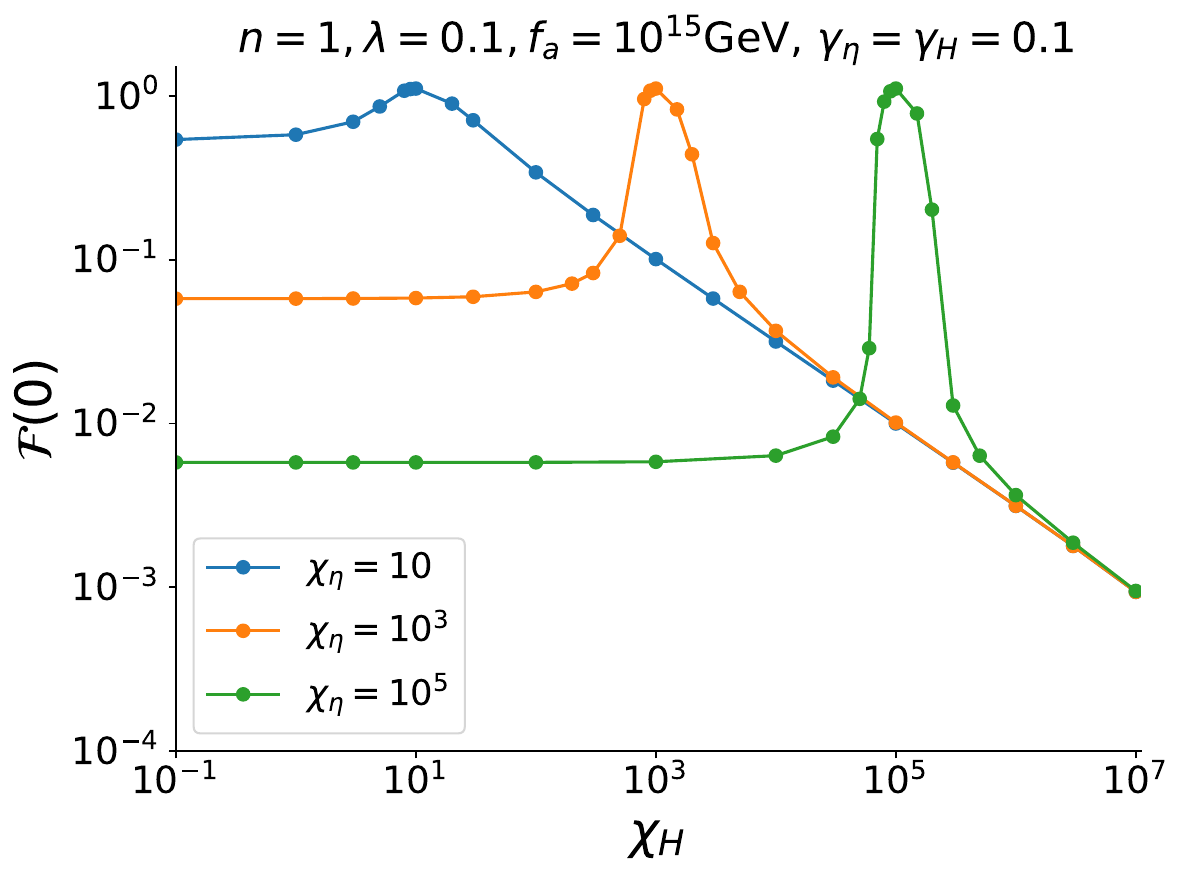}
  \hfill
  \includegraphics[clip, width=0.48\textwidth]{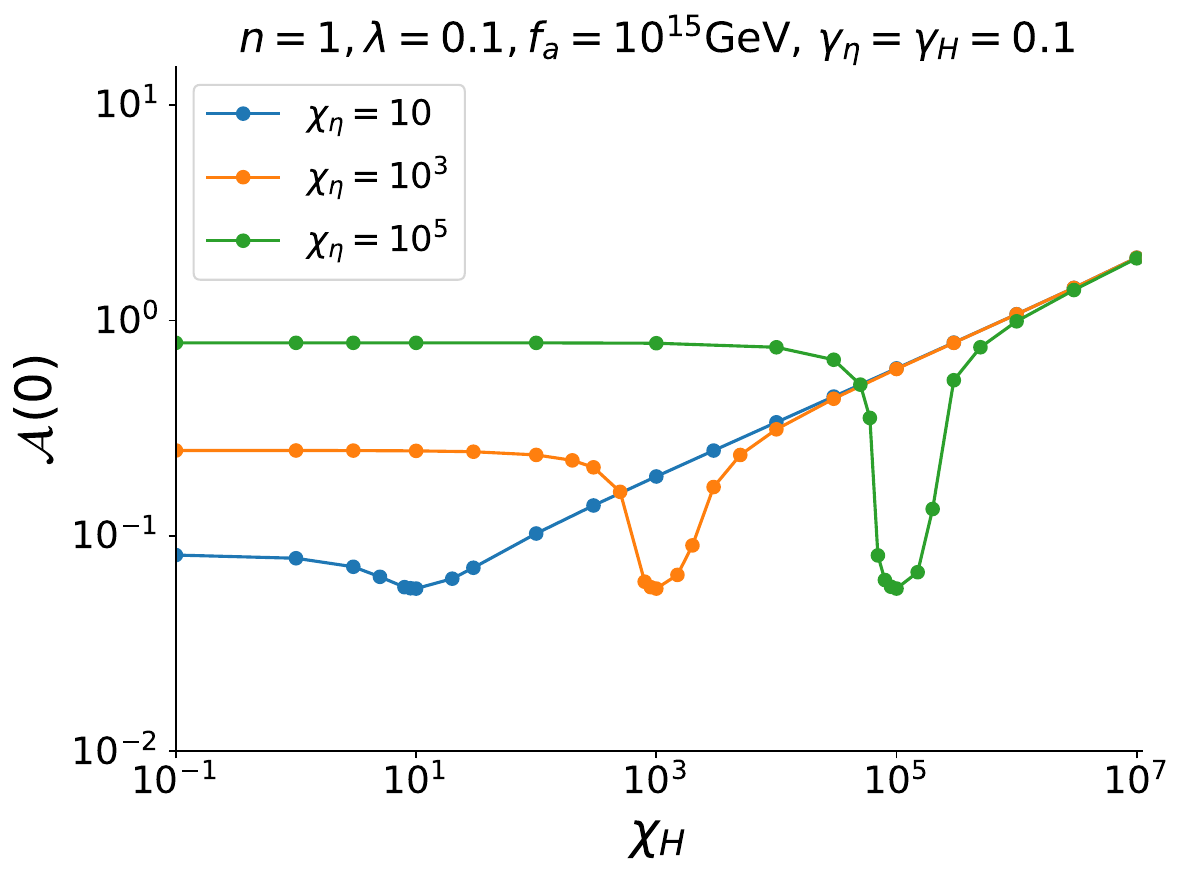}
  \caption{The behaviors of ${\mathcal F}(0)$ and ${\mathcal A}(0)$ in the case of the Nieh--Yan and Holst terms with $\gamma_\eta = \gamma_H = 0.1$.}
  \label{fig:F0A0_Nieh_Holst_01}
\end{figure}

Finally, we consider the combination of the Nieh--Yan and Holst terms.
Figs.~\ref{fig:F0A0_Nieh_Holst} and \ref{fig:F0A0_Nieh_Holst_01} show the behavior of $\mathcal{F}(0)$ and $\mathcal{A}(0)$ as functions of $\chi_H$ for various values of $\chi_\eta$, with $\gamma_H$ fixed to $10$ and $0.1$, respectively.
As shown in the figures, around $\chi_H \simeq \chi_\eta$, $\mathcal{F}(0)$ approaches $\mathcal{O}(1)$ and $\mathcal{A}(0)$ drops to $\sim 0.06$ because the factor $G$ in Eq.~\eqref{eq:G_NH} vanishes; this corresponds to the Palatini limit with $\xi=0$.
In other regions, the behavior depends on the relative magnitudes of $\chi_\eta$ and $\chi_H$.
In the large $\gamma_H$ case, the $\gamma_H^2$ factors in the numerator and the denominator of Eq.~\eqref{eq:G_NH} cancel, so that $G \sim 18 \mathcal{F}^2 (\chi_\eta - \chi_H)^2$; hence, for $\chi_\eta > \chi_H$, the factor $G$ mostly depends on $\chi_\eta$, and as $\chi_\eta$ increases, $\mathcal{F}(0)$ decreases and $\mathcal{A}(0)$ increases.
Conversely, for $\chi_\eta < \chi_H$, the factor $G$ primarily depends on $\chi_H$, and both $\mathcal{F}(0)$ and $\mathcal{A}(0)$ asymptotically converge to their respective universal values at large $\chi_H$ regardless of the initial difference in $\chi_\eta$.
For the small $\gamma_H$ case, $\mathcal{F}(0)$ and $\mathcal{A}(0)$ exhibit behavior similar to that in the large $\gamma_H$ case.
However, in this case, since $\gamma_H(\chi_\eta - \chi_H)$ in Eq.~\eqref{eq:G_NH} reduces to $\xi_\eta - \xi_H$ for $\gamma_\eta = \gamma_H$, the factor $G$ mostly depends on $\xi_\eta$ for $\chi_\eta > \chi_H$ and on $\xi_H$ for $\chi_\eta < \chi_H$.

\section{Inflation with non-minimal couplings in MAG}
\label{sec:inflation}

In this appendix, we summarize the inflationary analysis used in Sec.~\ref{sec:NA}.
The purpose of this analysis is to check whether the parameter regions required to solve the axion quality problem are compatible with inflationary observables.
We follow the numerical procedure and the formulation of inflation with non-minimal couplings to the curvature scalar, the Holst term, and the Nieh--Yan term developed in Refs.~\cite{Langvik:2020nrs,Shaposhnikov:2020gts}.

In the main text, the wormhole analysis is performed in the Jordan frame.
For the inflationary dynamics, however, it is useful to make contact with the Einstein frame description, where the gravitational sector takes the form of the Einstein--Hilbert action.
The relation between inflationary perturbations in the Jordan and Einstein frames has been studied since the early analyses of non-minimally coupled inflation~\cite{Makino:1991sg}.
More recently, the behavior of the comoving curvature perturbation under the conformal transformation was clarified in Refs.~\cite{Karciauskas:2022jzd,Diaz:2023tma}.
In general, the comoving curvature perturbation defined with respect to the effective fluid is not exactly conformally invariant.
However, for an effectively single-field slow-roll trajectory, the Jordan and Einstein frame descriptions agree on superhorizon scales up to gradient-suppressed corrections.
Therefore, at leading order in the slow-roll approximation, we evaluate the amplitude of the curvature perturbation $A_{\mathcal R}$, the scalar spectral index $n_s$, and the tensor-to-scalar ratio $r$ in the Friedmann--Lemaître--Robertson--Walker (FLRW) metric,
\begin{align}
    ds^2 = -dt^2 + a^2(t)(dx^2+dy^2+dz^2),
\end{align}
where $a(t)$ denotes the scale factor.
These observables are equivalent in the Jordan and Einstein frames at this order.

The reduction to an effectively single-field trajectory is justified as follows.
Owing to the global $\U(1)$ symmetry of the complex scalar field, the angular mode $\theta$ obeys a current-conservation equation in the homogeneous FLRW background,
\begin{equation}
    \frac{d}{dt}\left(a^3 f^2 \dot{\theta}\right)=0 .
\end{equation}
In the inflationary background, we work in the sector with vanishing homogeneous PQ charge.
The conserved quantity is then zero, and hence $\dot{\theta}=0$ is a consistent solution.
Even if a nonzero homogeneous charge were present, its energy density $\rho_\theta=\frac12 f^2\dot\theta^2$ would be rapidly diluted during inflation as $\rho_\theta\propto a^{-6}f^{-2}$.
Thus, the inflationary dynamics is effectively reduced to the radial single-field trajectory.
In the Einstein frame, after solving for the independent connection, the action of the radial component $f$ takes the form
\begin{equation}
    S_f
    =
    \int d^4x\sqrt{-g}
    \left[
        \frac{M^2_p}{2}R^{\diamond}
        -\frac{1}{2} K(f)\,
        g^{\mu\nu}\partial_\mu f\,\partial_\nu f
        -U(f)
    \right],
\end{equation}
where the effects of the non-minimal couplings to the curvature scalar, the Holst term, and the Nieh--Yan term are encoded in the non-canonical kinetic factor $K(f)$.
Introducing the canonical inflaton field $\varphi$ by
\begin{equation}
    \frac{d\varphi}{df}=\sqrt{K(f)},
\end{equation}
the inflationary dynamics reduces to the standard single-field slow-roll system with the canonical potential defined by $\mathcal{U}(\varphi)\equiv U(f(\varphi))$.
In the slow-roll regime, $A_{\mathcal R}$, $n_s$, and $r$ can be expressed in terms of the slow-roll parameters, $\epsilon_{\mathcal{U}}=\frac{M^2_p}{2}\left(\frac{\mathcal{U}_\varphi}{\mathcal{U}} \right)^2, ~\eta_{\mathcal{U}}=M^2_p\frac{\mathcal{U}_{\varphi\varphi}}{\mathcal{U}}$.
Then, we can obtain the values of the canonical field $\varphi$ at which the primordial scalar perturbation is generated~$(\varphi_{\ast})$ and at which inflation ends~$(\varphi_{\text{end}})$ as
\begin{align}
    A_{\mathcal R}(\varphi_{\ast})&=A^{\ast}_{\mathcal R},\\
    \varphi_{\text{end}}&\equiv \max(\varphi:\epsilon_{\mathcal{U}}(\varphi)=1\text{ or }|\eta_{\mathcal{U}}(\varphi)|=1).
\end{align}
Finally, we numerically calculate the number of e-folds, $N_e=\log \left(\frac{a_{\text{end}}}{a_{\ast}} \right)$, using these field values.
$a_\ast$ and $a_{\text{end}}$ are the scale factors evaluated at $\varphi_\ast$ and $\varphi_{\text{end}}$, respectively.
Using a Markov chain Monte Carlo sampler \cite{Foreman-Mackey:2012any}, our analysis identifies the parameter regions where a fixed value of $N_e$ is realized.

For the observational constraints, we adopt the following ACT-compatible values and ranges \cite{BICEP:2021xfz, AtacamaCosmologyTelescope:2025blo, AtacamaCosmologyTelescope:2025nti},
\begin{align}
    A^{\ast}_{\mathcal R} &\simeq 2.1\times 10^{-9},\\
    n_s &= 0.9743 \pm 0.0034,
    \qquad
    r < 0.038 ,
\end{align}
evaluated at the pivot scale $k_\ast = 0.05\,{\rm Mpc}^{-1}$.
In the main text, we adopt the following benchmark conditions,
\begin{equation}
    N_e = 80,
    \qquad
    0.97 < n_s < 0.98,
    \qquad
    r < 0.038 .
\end{equation}
Since a somewhat larger number of e-folds is required in the parameter regions considered here to achieve compatibility with the ACT-preferred range of $n_s$, we use $N_e=80$ as a representative benchmark.

\begin{figure}[h]
  \centering
  \includegraphics[clip, width=0.48\textwidth]{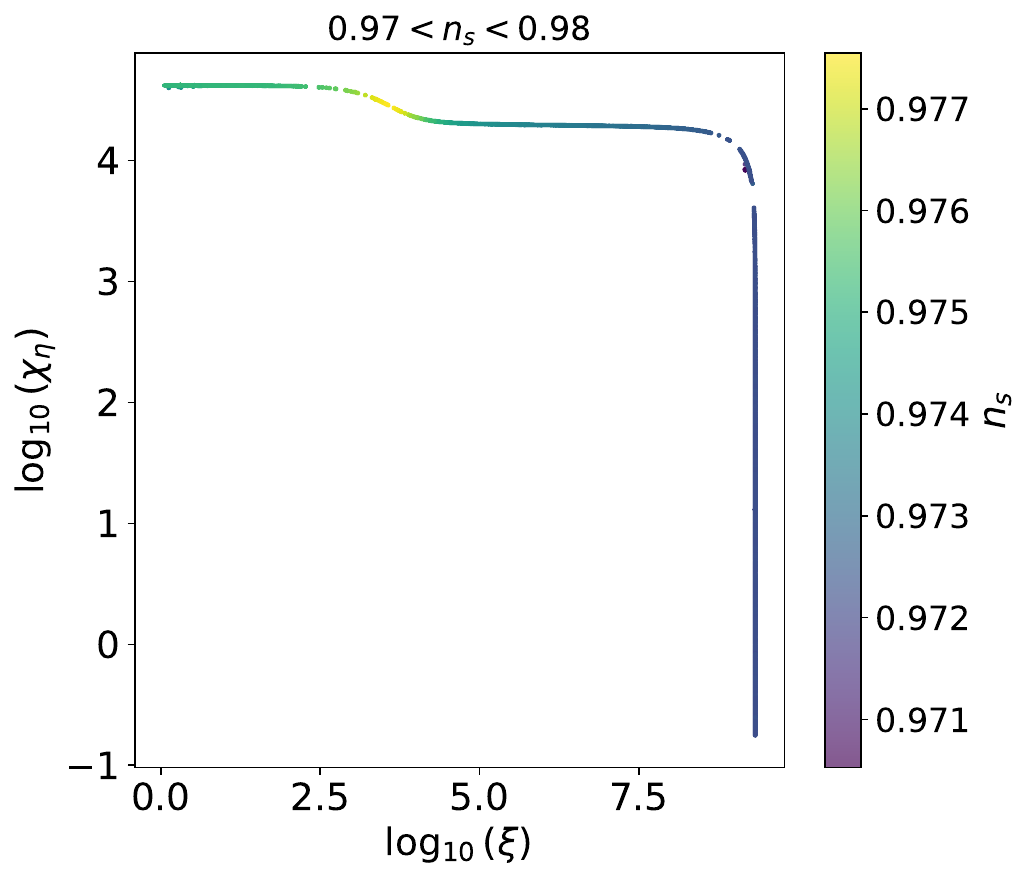}
  \hfill
  \includegraphics[clip, width=0.48\textwidth]{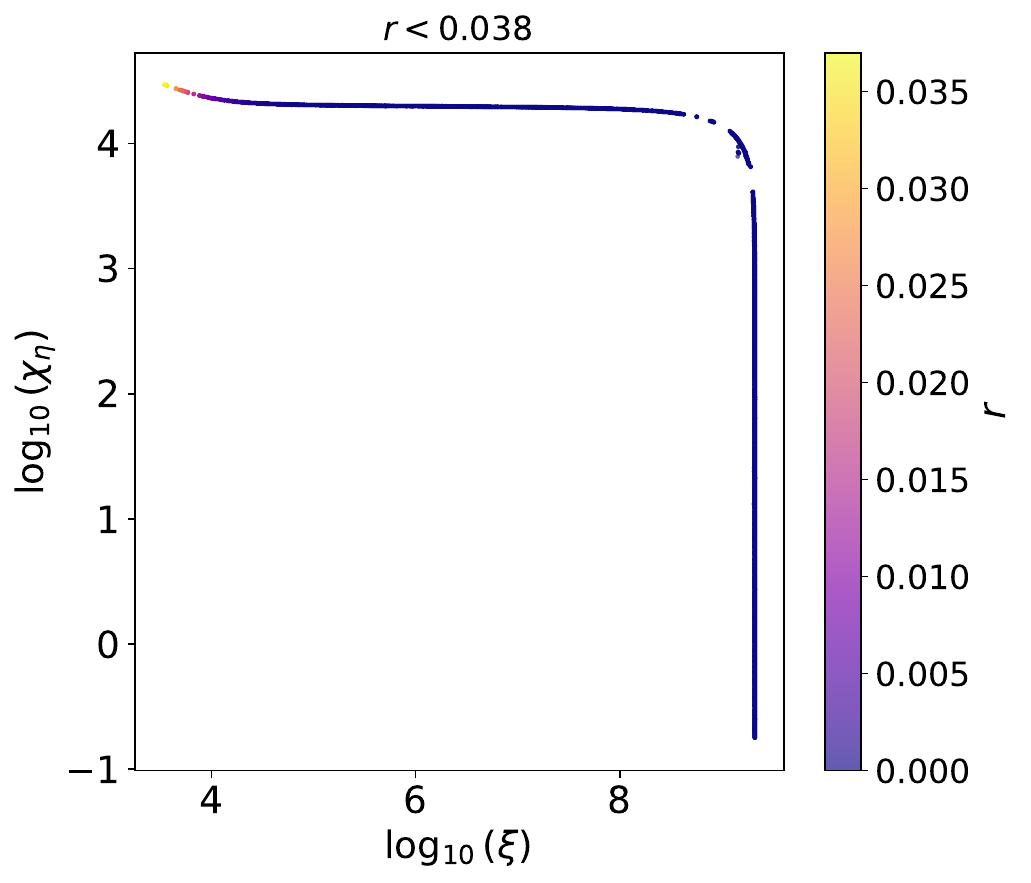}
  \caption{The spectral index $n_s$ and tensor-to-scalar ratio $r$ in the case of the non-minimal couplings to the curvature scalar and the Nieh--Yan term.
  The benchmark observational conditions are $0.97 < n_s < 0.98$ and $r < 0.038$.}
  \label{fig:InflationXiNieh}
\end{figure}

\begin{figure}[h]
  \centering
  \includegraphics[clip, width=0.48\textwidth]{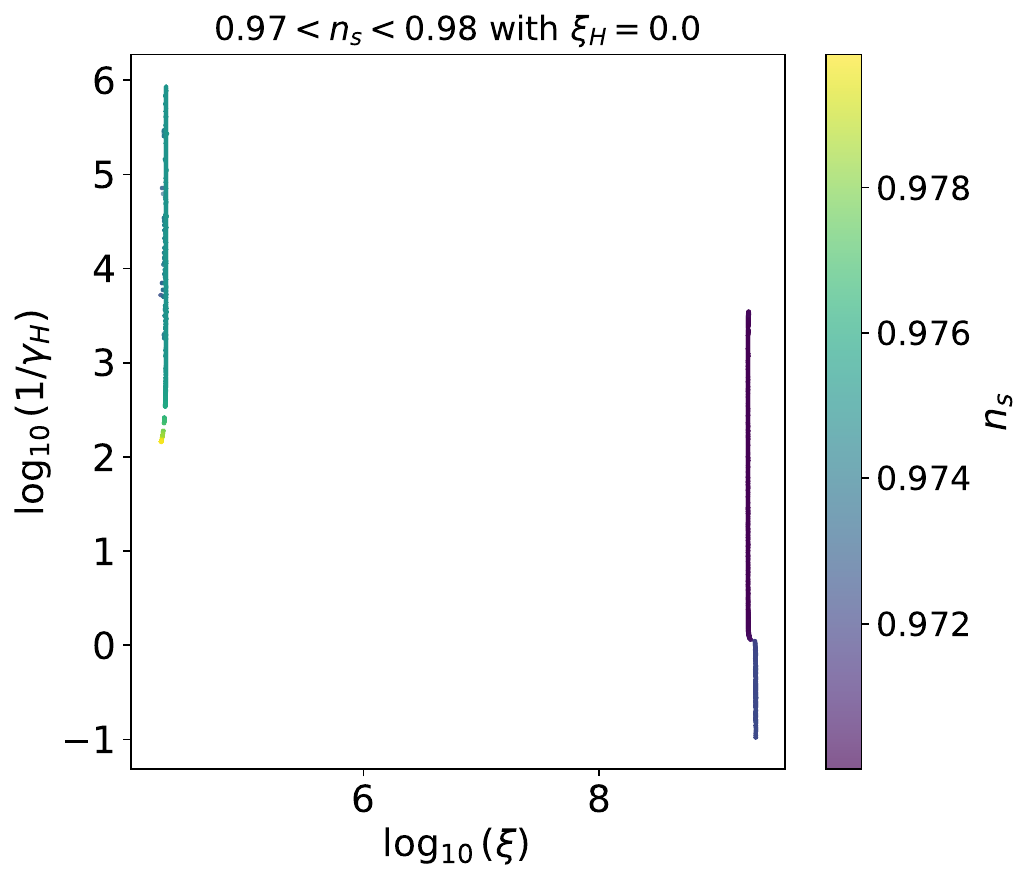}
  \hfill
  \includegraphics[clip, width=0.48\textwidth]{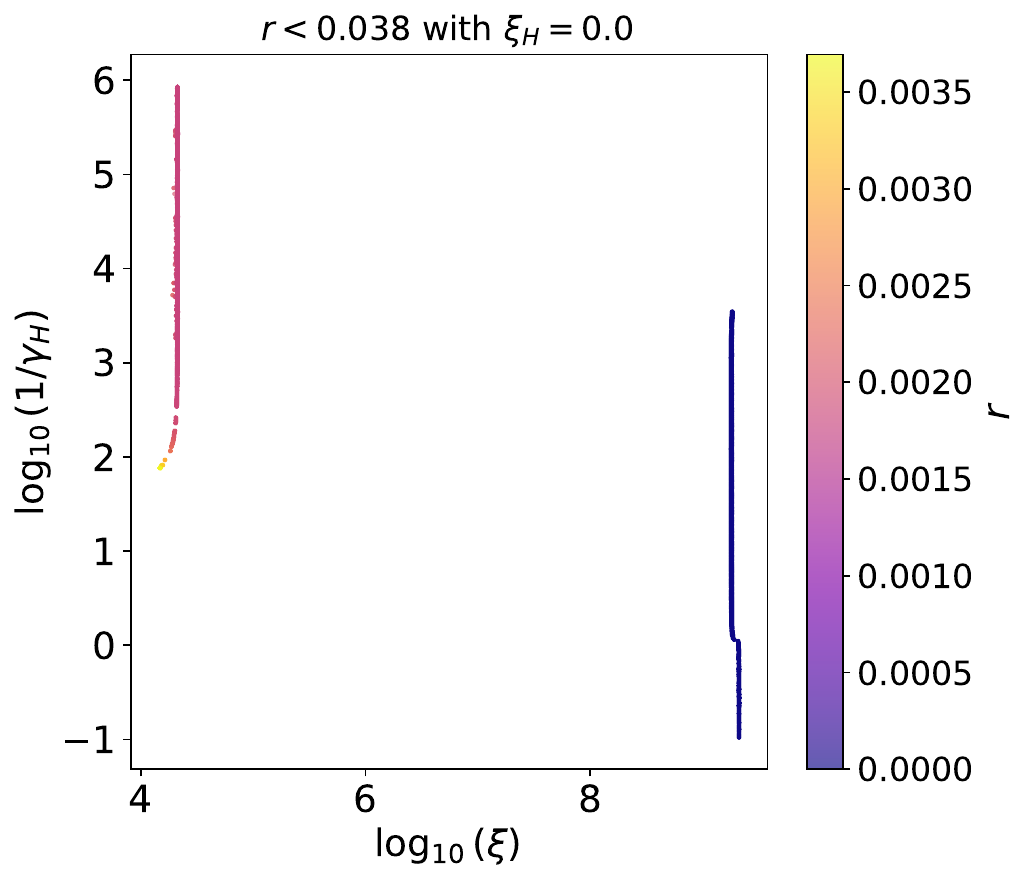}
  \caption{The spectral index $n_s$ and tensor-to-scalar ratio $r$ in the case of the non-minimal couplings to the curvature scalar and the Barbero--Immirzi parameter.
  The benchmark observational conditions are $0.97 < n_s < 0.98$ and $r < 0.038$.}
  \label{fig:InflationXiGamma}
\end{figure}

\begin{figure}[h]
  \centering
  \includegraphics[clip, width=0.48\textwidth]{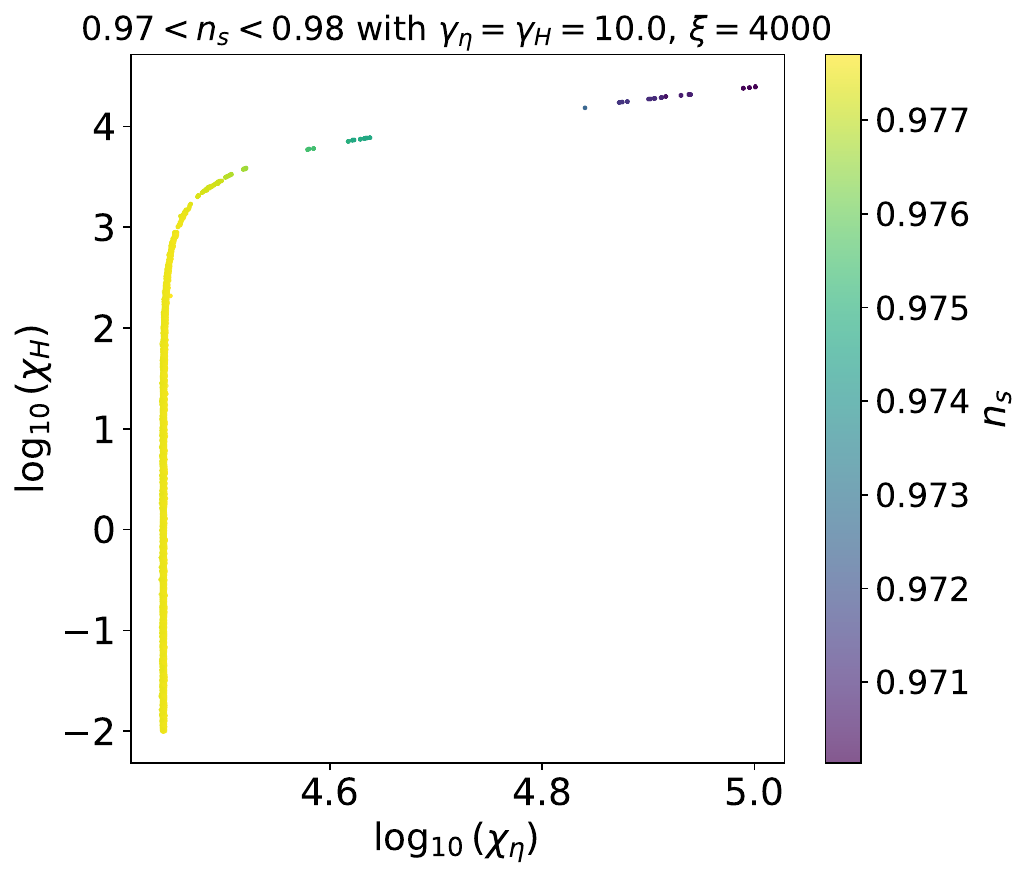}
  \hfill
  \includegraphics[clip, width=0.48\textwidth]{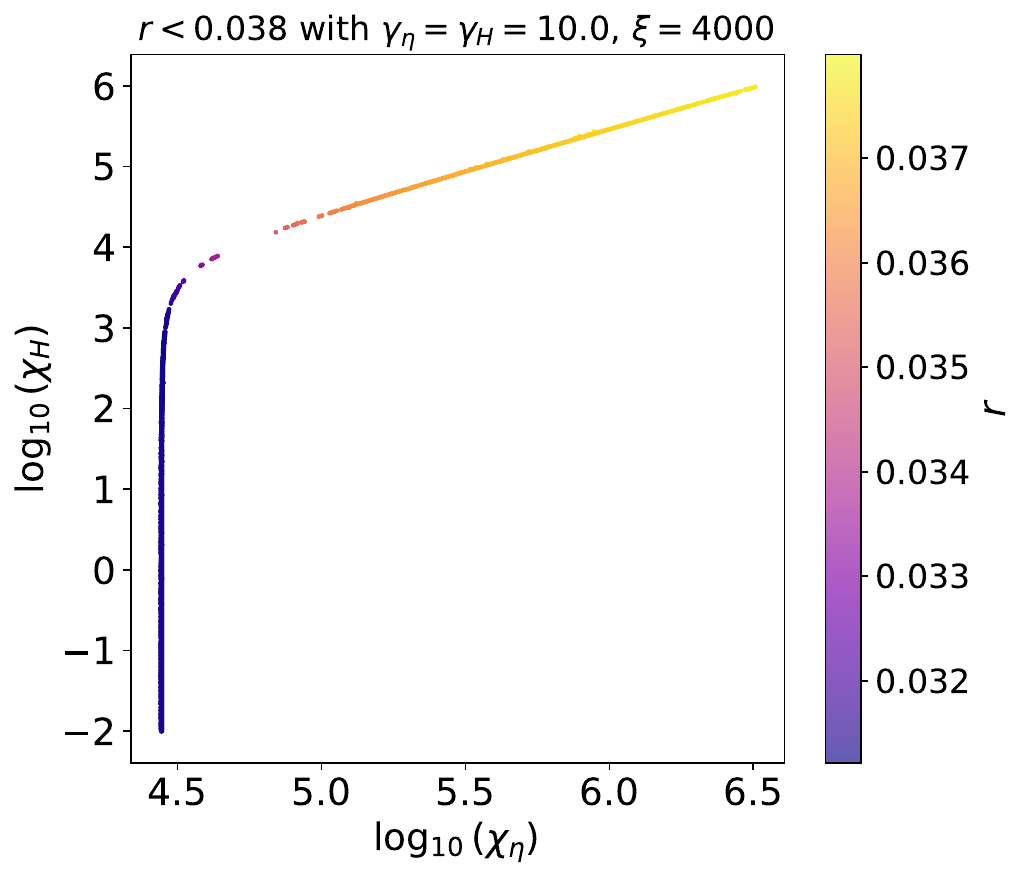}
  \caption{The spectral index $n_s$ and tensor-to-scalar ratio $r$ in the case of the non-minimal couplings to the Nieh--Yan and Holst terms with $\xi=4000$.
  The benchmark observational conditions are $0.97 < n_s < 0.98$ and $r < 0.038$.}
  \label{fig:InflationNiehHolst}
\end{figure}

\begin{figure}[h]
  \centering
  \includegraphics[width=0.6\textwidth]{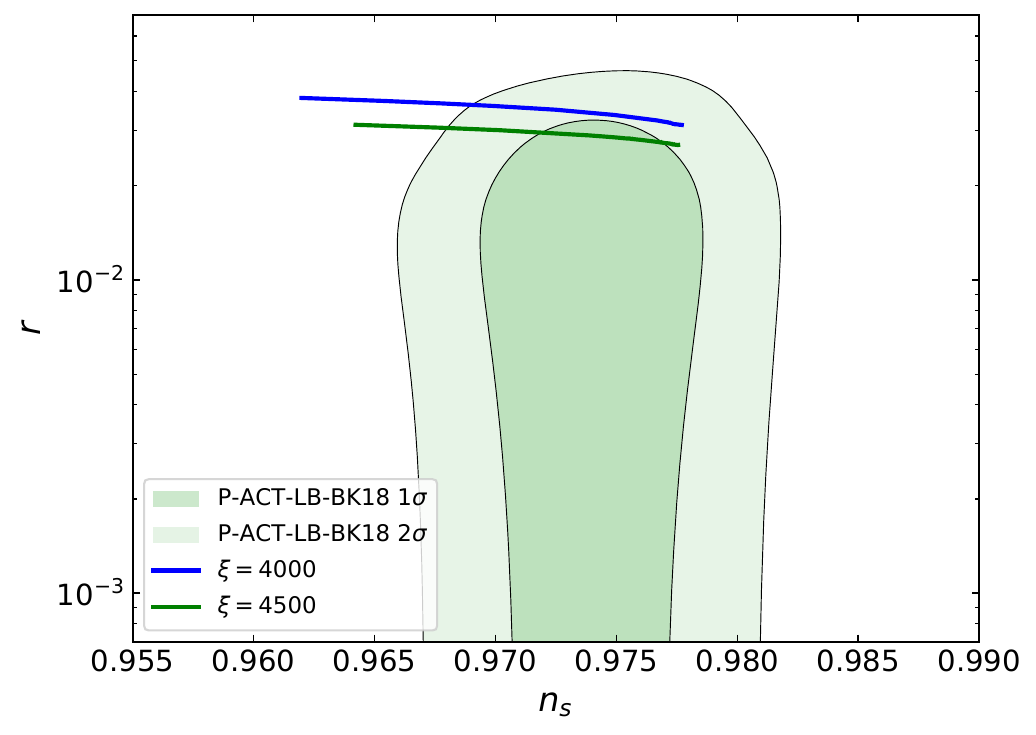}
\caption{
Predictions in the $(n_s,r)$ plane for representative values $\xi=4000$ and $4500$ at $N_e=80$.
The dark and light purple regions show the marginalized $68\%$ and $95\%$ credible regions, respectively, obtained from the public P-ACT-LB-BK18 chain of Ref.~\cite{AtacamaCosmologyTelescope:2025nti} using \textsc{GetDist}~\cite{Lewis:2019xzd}.
This data combination includes Planck 2018 and ACT DR6 primary CMB anisotropies~\cite{AtacamaCosmologyTelescope:2025blo}, ACT DR6 and Planck NPIPE CMB lensing, DESI Year-1 BAO, and BK18 B-mode data~\cite{BICEP:2021xfz}.
The tensor-to-scalar ratio is defined at the pivot scale $k_*=0.05\,{\rm Mpc}^{-1}$.
The colored curves show the predictions of the model with non-minimal couplings to the Nieh--Yan and Holst terms and finite $\xi$.
}
\label{fig:OverlayNiehHolst}
\end{figure}

Figs.~\ref{fig:InflationXiNieh}, \ref{fig:InflationXiGamma}, and \ref{fig:InflationNiehHolst} show the inflationary observables in the parameter regions used in Sec.~\ref{sec:NA}.
While the wormhole analysis is conservatively restricted to $\xi \leq M_p^2/f_a^2$ to ensure $M^2_p+\xi(f^2 - f_{a}^2)>0$ for arbitrary field configurations, the inflationary analysis is extended beyond this range because $M^2_p+\xi(f^2 - f_{a}^2)$ remains positive along the inflationary trajectory.
Fig.~\ref{fig:InflationXiNieh} shows the spectral index and the tensor-to-scalar ratio in the case of the curvature scalar and the Nieh--Yan term.
Fig.~\ref{fig:InflationXiGamma} shows the corresponding result for the curvature scalar coupling combined with the Barbero--Immirzi parameter.
Finally, Fig.~\ref{fig:InflationNiehHolst} shows the result in the case of the Nieh--Yan and Holst terms with a finite curvature scalar coupling.
In the last case, increasing $\xi$ moves the predicted values of $(n_s,r)$ toward the observationally favored region as shown in Fig.~\ref{fig:OverlayNiehHolst}.
This motivates the representative choice $\xi=4000$ used in Figs.~\ref{fig:WormInfNiehHolstXi4000} and \ref{fig:InflationNiehHolst}.

\clearpage

\bibliography{Worm_MAG}
\bibliographystyle{JHEP}
\end{document}